\documentclass[%
 reprint,
nofootinbib, 
 amsmath,amssymb,
 aps,
]{revtex4-2}

\usepackage{graphicx}
\usepackage{dcolumn}
\usepackage{bm}

\usepackage{braket}
\usepackage{tikz}
\usetikzlibrary{quantikz2}
\usepackage{subcaption}
\usepackage{amsmath}
\usepackage{mathrsfs}
\usepackage{subcaption}
\usepackage{ragged2e}
\usepackage{gensymb}
\usepackage{comment}
\usepackage{hyperref}
\usepackage{cleveref}
\usepackage{soul}

\begin{document}


\title{Quantum circuits for simulating neutrino propagation in matter}

\author{Sandeep Joshi}
 \email{sjoshi@barc.gov.in}
\author{Garima Rajpoot}
\author{Prashant Shukla}

\affiliation{%
TNP\&QC Section, Nuclear Physics Division, Bhabha Atomic Research Centre, Mumbai 400085, India
}%





\begin{abstract}
Quantum simulation of particle phenomena is a rapidly advancing field of research. With the widespread availability of quantum simulators, a given quantum system can be simulated in numerous ways, offering flexibility in implementation and exploration. Here, we perform quantum simulation of neutrino propagation in matter, a phenomenon that plays a crucial role in neutrino oscillations.  We present quantum circuits with novel gate arrangements to simulate neutrino propagation in both constant and varying matter density profiles. The oscillation probabilities are determined by encoding and measuring the qubit states in the neutrino flavor basis, showing excellent agreement with theoretical predictions.
\end{abstract}


\maketitle


\section{\label{sec:1}Introduction}
Neutrino oscillations is still the only observed phenomenon that does not fit into the standard model of particle physics. While numerous experiments are running worldwide to further investigate this phenomenon, it remains a highly explored topic from a theoretical perspective. Neutrino oscillation is a quantum mechanical phenomenon that arises from the superposition of the three neutrino flavor states. Due to tiny neutrino mass, this superposition persists over macroscopic distance scales, resulting in oscillations between neutrino flavors. This has been observed in multiple experiments across different parameter regimes \cite{Fukuda:1998mi, Ahmad:2002jz, Araki:2004mb}. 

As neutrinos propagate through matter, they undergo coherent forward scattering with the particles in the medium. This induces a flavor-dependent effective mass for the neutrinos and modifies the neutrino flavor mixing. The density profile of the medium plays an important role in determining the oscillation probability. In a medium with a periodic density profile such as the Earth \cite{Giunti:2007ry}, or slowly varying density such as the Sun \cite{Joshi:2019dcj}, resonances can occur which lead to an enhanced probability of flavor conversion \cite{Kuo:1989qe}.

Quantum mechanically, neutrino propagation can be described by the evolution of the neutrino flavor states in the system Hilbert space. For two flavor neutrino oscillations, the Hilbert space is a Bloch sphere $S^2$. The evolution of neutrino flavor states can be geometrically represented as precession of the spin-polarization vector in the presence of an effective magnetic field \cite{Joshi:2020djx}.  A qubit with its two-dimensional Hilbert space naturally provides a way to encode the two neutrino flavor states. The time evolution of neutrino flavor states can then be mimicked by applying quantum gates on the encoded qubit. Measuring the qubit states would then give us the probability of neutrino flavor oscillations. In this way, the neutrino oscillation phenomenon can be simulated in a quantum simulator or a quantum processing unit (QPU). 

More generally, quantum simulation refers to the simulation of a quantum system using another controllable quantum system that can be realized in a laboratory \cite{Feynman:1981tf, Lloyd:1996aai, Georgescu:2013oza}. The dynamics of the controlled system are carefully engineered to achieve the desired unitary evolution, which is implemented through a quantum circuit composed of a sequence of quantum gates. This approach, known as digital quantum simulation, is agnostic to the specific hardware used to perform the simulation. A quantum simulator typically refers to a software-based tool that emulates quantum circuits on classical hardware, allowing for efficient testing of algorithms without any  hardware constraints \cite{Johnson:2014igy}. In contrast, a QPU consists of actual quantum hardware, such as superconducting qubits or trapped ions, where computations are carried out physically. 

With the availability of publicly accessible QPUs and quantum simulators, such as the one provided by IBM \footnote{\url{https://quantum.ibm.com/}, \newline \url{https://qiskit.github.io/qiskit-aer/stubs/qiskit_aer.AerSimulator.html}}, simulating nuclear and particle physics phenomena has generated significant interest \cite{Garcia-Ramos:2023rtd, Roggero:2019myu, singh2025quantum, Chen:2021cyw, Bauer:2022hpo, DiMeglio:2023nsa, Bauer:2023qgm}. In the context of neutrinos, several studies have been performed to simulate the neutrino oscillation phenomenon using various hardware platforms and quantum gate implementations \cite{noh2012quantum, Arguelles:2019phs, Jha:2021itm, Nguyen:2022snr, Singh:2024vpu, Rajpoot:2025nzu}. For example, in \cite{Arguelles:2019phs}, two and three flavor vacuum neutrino oscillations were simulated using the single and two qubit gates in the Qiskit simulator and the IBM QPU. In \cite{Nguyen:2022snr}, the three transmon \cite{Koch:2007hay} levels were used to encode the neutrino flavors, and the oscillations were simulated using qutrit gates. In addition to the superconducting qubits, other quantum computing hardware, such as nuclear magnetic resonance (NMR) processor \cite{Singh:2024vpu} and trapped ion technology \cite{noh2012quantum}, have been used to simulate neutrino oscillations. In astrophysical environments with high neutrino density, such as core-collapse supernovae, quantum simulation of collective neutrino oscillations  \cite{Hall:2021rbv, Yeter-Aydeniz:2021olz, Illa:2022zgu, Amitrano:2022yyn, Siwach:2023wzy, Turro:2024shh, Spagnoli:2025etu} is expected to offer valuable insights into the role of quantum many-body effects in neutrino dynamics.

The current quantum hardware faces several limitations such as small number of qubits, restricted connectivity between qubits and noisy gate operations. These so called noisy intermediate-scale quantum (NISQ) devices \cite{Preskill:2018jim} demand algorithms with low circuit depth and fast gate execution times. Thus, while designing quantum circuits it is important to select gate sets and circuit architectures that maximize hardware performance. Several techniques have been proposed for such circuit optimizations, typically involving modifications to the gate structure during circuit design. For instance, \cite{Vatan:2004nmz, Shende:2004gqq} present quantum circuits with minimum number of \texttt{CNOT} gates required to implement an arbitrary two-qubit unitary. The work in  \cite{Jang:2021ary} identifies low amplitude basis states and removes redundant gates to reduce circuit complexity. References \cite{Zulehner:2019lkn, Wille:2019lxk, Lin:2022cyg} develop efficient methods to map quantum circuits onto specific backend hardware, thereby reducing gate overhead. In \cite{DeCross:2022kuu},  the use of mid-circuit measurements and qubit reuse enables the compilation of circuits that require fewer physical qubits. In \cite{Fischer:2022dbr, Kiktenko:2023ytz}, qudit gates having Hilbert space dimension $d>2$ are employed to reduce the circuit depth. Reference \cite{Vezvaee:2024ywq} demonstrates that symmetric compilation of virtual-$Z$ (\texttt{VZ}) gates \cite{McKay:2017rej} leads to an improved gate fidelity in open quantum system settings. In particular, the use of \texttt{VZ} gates has become a standard optimization technique in quantum circuit compilation, especially for NISQ-era devices. These gates are implemented through frame updates at the software level, requiring no physical pulses, and thus contribute neither to circuit duration nor to gate error.

In this work, we present quantum circuits to simulate two flavor neutrino propagation in matter. Section \ref{sec:2} presents the theoretical details where we describe the expressions of oscillation probability in both constant and varying density profiles. In Section \ref{sec:3}, we discuss the encoding of the neutrino propagator in terms of quantum gates. We describe the gates, which can be implemented both physically on hardware and virtually through software. Quantum circuits incorporating both types of gates are presented for simulating neutrino propagation through Earth.  We also perform a benchmarking comparison to demonstrate that our approach leads to a reduction in both circuit depth and execution time. In Section \ref{sec:4}, we present an effective unitary that can be used to simulate solar neutrino propagation. This unitary is implemented using two qubit entangling gates. All the above circuits are first simulated using Qiskit's AerSimulator package\footnote{Qiskit AerSimulator builds a noise model based on the calibration data from the QPU (\texttt{ibm\_brisbane} in our case), enabling realistic quantum simulations.} and subsequently executed on \texttt{ibm\_brisbane}, a cloud-based IBM quantum processor. In Section \ref{sec:5} we present conclusions and discussions. 

\section{\label{sec:2} Neutrino propagation in matter}
Neutrino flavor mixing is characterized by a unitary mixing matrix $\mathscr{U}$ such that $\ket{\nu_\alpha}= \sum_i \mathscr{U}_{\alpha i} \ket{\nu_i}$, where $\alpha = e, \mu ~ \mbox{or}~ \tau$ represents the neutrino flavor state and $i= 1, 2, 3$ represents the mass eigenstate with mass $m_i$. We restrict ourselves to the case of two flavor neutrino  oscillations.  The unitary matrix $\mathscr{U}$  in this case is parametrized by an angle $\theta$, called mixing angle:
\begin{equation} \label{unit-vac}
    \mathscr{U}(\theta) = \begin{pmatrix}
        \cos \theta & \sin \theta \\
        - \sin \theta & \cos \theta
    \end{pmatrix}.
\end{equation}

The propagation of neutrinos in background matter induces an effective potential due to the coherent forward scattering of neutrinos with the electrons in the medium. The evolution of neutrino flavor states can be described by the equation
\begin{align} \label{matter-evol}
     i \frac{d}{dt} \ket{\psi(t)} = H_{\rm F} \ket{\psi(t)}, 
    \end{align}
where $\ket{\psi(t)} = (\nu_e(t)~ \nu_\mu(t))^{\rm T}$, $\nu_e(t)$ and $\nu_\mu(t)$ being the probability amplitudes for neutrino to be in state $\ket{\nu_e}$ and $\ket{\nu_\mu}$, respectively,  and $H_{\rm F}$ is the Hamiltonian in the flavor basis given by \cite{Kuo:1989qe}
    \begin{equation} \label{matter-evol-2}
    H_F =  \frac{1}{2E} \Big[\mathscr{U} (\theta) \begin{pmatrix}
        -\Delta m^2/2 & 0 \\
        0 & \Delta m^2/2
    \end{pmatrix} \mathscr{U}^\dagger(\theta)  + \begin{pmatrix}
        A & 0 \\
        0 & 0
    \end{pmatrix} \Big],
\end{equation}
where $E$ is the neutrino energy and $\Delta m^2= m_2^2- m_1^2$. The parameter $A$ characterizes the effect of matter on neutrino propagation and is given by $A= 2 \sqrt{2} G_{\rm F} Y_e \rho E/m_n$, where $G_{\rm F}$ is the Fermi constant, $\rho$ is density of matter, $Y_e$ is the number of electrons per nucleon, and $m_n$ is the nucleon mass. The Hamiltonian $H_{\rm F}$ can be diagonalized by the unitary matrix \eqref{unit-vac} with parameter $\theta$ replaced by an effective mixing angle in matter $\theta_m$ given by 
\begin{equation} \label{mixing-angle}
    \sin 2\theta_m=    \frac{\sin 2 \theta}{\sqrt{(\cos 2 \theta- \beta)^2+ \sin^2 2 \theta}},
\end{equation}
where $\beta = A/\Delta m^2$. The eigenvalues of the diagonalized Hamiltonian give the mass-squared difference in matter:
\begin{equation}\label{mass-squared}
    \Delta m^2_m = \Delta m^2 \sqrt{(\cos 2 \theta- \beta)^2+ \sin^2 2 \theta}.
\end{equation}
In a medium with constant density, solving Eq. \eqref{matter-evol} gives us the evolution vector
\begin{equation}
    \ket{\psi(t)} = \mathscr{U}(\theta_m) \mathcal{U}(\phi) \mathscr{U}^\dagger(\theta_m) \ket{\psi(0)},
\end{equation}
where  $\phi= \Delta m_m^2 t/2E$ and 
\begin{equation}
    \mathcal{U}(\phi) = \begin{pmatrix}
            e^{-i \phi/2} & 0 \\
            0 & e^{i \phi/2}
    \end{pmatrix}.
\end{equation}

Since neutrinos propagate nearly at the speed of light, we can make the replacement $t \rightarrow x$, where $x$ is the distance traveled. If neutrinos are produced in the flavor state $\ket{\nu_\mu}$, and are detected after traveling a certain distance from the source, then the oscillation probability is given by
\begin{align} \label{prob-1}
    P(\nu_\mu \rightarrow \nu_e) =& \big|\bra{\nu_e}\mathscr{U}(\theta_m) \mathcal{U}(\phi) \mathscr{U}^\dagger(\theta_m)\ket{\nu_\mu}\big|^2 \nonumber \\
    =& \Bigg|\begin{pmatrix}
        1 & 0
    \end{pmatrix} \begin{pmatrix}
        \cos \theta_m & \sin \theta_m \\
        - \sin \theta_m & \cos \theta_m
    \end{pmatrix} \begin{pmatrix}
        e^{-i \phi/2} & 0 \\
            0 & e^{i \phi/2}
    \end{pmatrix} \nonumber \\& \begin{pmatrix}
        \cos \theta_m & -\sin \theta_m \\
         \sin \theta_m & \cos \theta_m
    \end{pmatrix}\begin{pmatrix}
        0 \\1 
    \end{pmatrix} \Bigg|^2.
\end{align}

\begin{figure}
\centering
\begin{subfigure}[b]{0.5\textwidth}
\centering
   \includegraphics[width=1\linewidth]{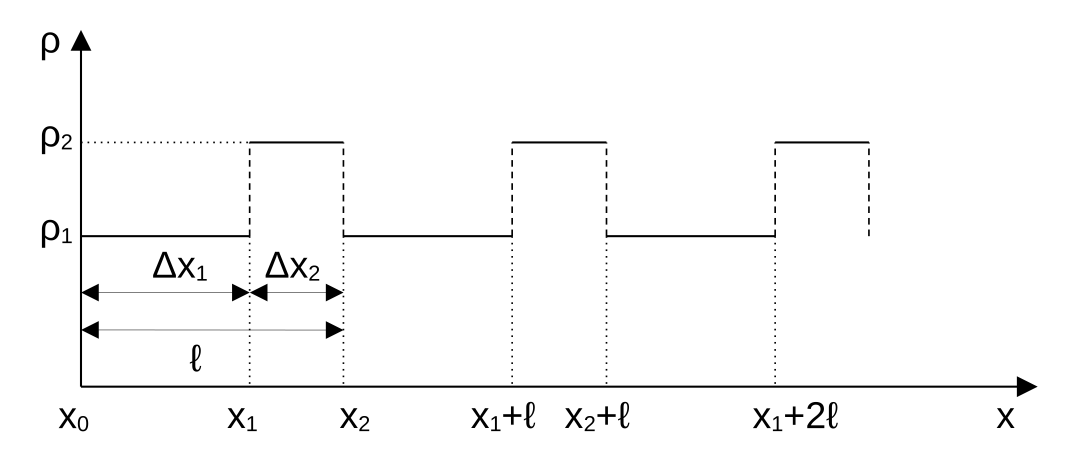}
   \caption{}
   \label{fig:slab} 
\end{subfigure}

\begin{subfigure}[b]{0.3\textwidth}
\centering
   \includegraphics[width=1\linewidth]{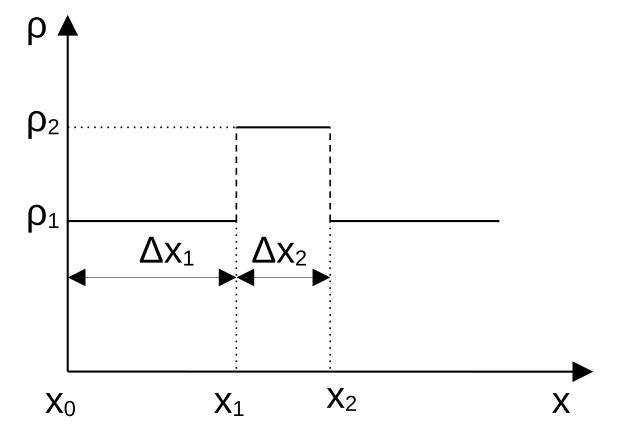}
   \caption{}
   \label{fig:earth}
\end{subfigure}

\caption[]{\RaggedRight (a) Slab profile for density profile of a medium. The medium consists of slabs with constant density $\rho_1$ and $\rho_2$ having lengths $\Delta x_1$ and $\Delta x_2$, respectively. Overall, there are $N$ slabs, and the medium has a periodicity $\ell$. (b) The density of Earth can be modeled using the slab profile with one and a half period. For the mantle region we set $\rho_1= 5$ g/cm$^3$ and $\Delta x_1= 5000$ km, and the core region has $\rho_2= 10$ g/cm$^3$ and $\Delta x_2= 2500$ km.}
\end{figure}

A useful model of neutrino propagation assumes a medium having a density that is piecewise constant and is periodic with a period $\ell$, as shown in Fig. \ref{fig:slab}. This is known as slab approximation \cite{Akhmedov:1999ty, Giunti:2007ry}. If the overall medium consists of  $N$ such slabs, then the time evolution vector for neutrino propagation is given by
\begin{align} \label{slab}
    \ket{\psi(x_N)}=& [\mathscr{U}(\theta_{m2}) \mathcal{U}(\phi_2) \mathscr{U}^\dagger(\theta_{m2})]_{(N)} ...[\mathscr{U}(\theta_{m2}) \mathcal{U}(\phi_2) \nonumber \\& \mathscr{U}(\theta_{m2})]_{(2)}[\mathscr{U}(\theta_{m1}) \mathcal{U}(\phi_1) \mathscr{U}^\dagger(\theta_{m1})]_{(1)} \ket{\psi(0)},
\end{align}
where  $\Delta m^2_{mk}$ and $\theta_{mk}~ (k= 1,2)$ are the effective mass-squared differences and mixing angles in the region with density $\rho_k$ respectively, and $\phi_k = \Delta m^2_{mk} \Delta x_k/2E$.
The periodic density profile can result in parametric enhancement of flavor transitions\cite{Giunti:2007ry}. This density model, with one and a half period of constant density slabs as shown in Figure \ref{fig:earth}, also serves as a good approximation of Earth's density \cite{Akhmedov:1998ui, Giunti:2007ry}.

As neutrinos propagate through a medium, another effect becomes significant when the density of the medium varies with distance. This happens when the resonance condition $\beta = \cos 2 \theta$ is satisfied, which results in enhancement of the mixing angle \eqref{mixing-angle}. For example, solar neutrinos are produced near the center of the Sun in regions with high density. If their energy is sufficiently high, they will eventually cross this resonance point as they move outward to low-density regions. If the density is a slowly varying function of the distance $x$ traveled by the neutrino, then the propagation is nearly adiabatic. The survival probability of $\nu_e$ as it propagates to vacuum is given by \cite{Kuo:1989qe}
\begin{align} \label{prob-2}
    P(\nu_e \rightarrow \nu_e) =& \begin{pmatrix}
        1 & 0
    \end{pmatrix} \begin{pmatrix}
        \cos^2 \theta & \sin^2 \theta \\
        \sin^2 \theta & \cos^2 \theta
    \end{pmatrix} \begin{pmatrix}
            1 & 0 \\
            0 & 1
    \end{pmatrix} \nonumber \\ & \begin{pmatrix}
        \cos^2 \theta_m & \sin^2 \theta_m \\
        \sin^2 \theta_m & \cos^2 \theta_m
    \end{pmatrix}  \begin{pmatrix}
        1 \\ 0
    \end{pmatrix} \nonumber \\
    = & \frac{1}{2}\big(1+ \cos 2 \theta \cos 2 \theta_m\big).
\end{align}
Here, the phase information is effectively lost, and instead of adding amplitudes as in Eq. \eqref{prob-1}, we add the probabilities of neutrino to be in the two mass eigenstates. This is known as MSW (Mikheyev-Smirnov-
Wolfenstein) effect \cite{Maltoni:2015kca}.

\section{\label{sec:3} Simulating matter effects using \texttt{VZ} gates}
The quantum simulation of two-flavor neutrino oscillations can be implemented using a single qubit, where the neutrino flavor states are mapped onto the computational basis states:
\begin{equation}
    \ket{\nu_e} \equiv \ket{0}=  \begin{pmatrix}
        1 \\ 0 
    \end{pmatrix}, ~~ \ket{\nu_\mu} \equiv \ket{1} =  \begin{pmatrix}
        0 \\1 
    \end{pmatrix}.
\end{equation}
A generic single qubit gate with three rotation angles that operates on a qubit state is given by the unitary
\begin{equation}\label{ugate}
    U(\Theta, \Phi, \lambda) = \begin{pmatrix}
        \cos \Theta/2 & - e^{i \lambda} \sin \Theta/2 \\
        e^{i \Phi} \sin \Theta/2 & e^{i (\Phi+ \lambda)} \cos \Theta/2 
    \end{pmatrix}.
\end{equation}
Geometrically, we can represent a qubit state as a vector from the center to the surface of the Bloch sphere and a unitary gate as rotations of this vector around the three axes of the sphere. For example, the gate \eqref{ugate} can be decomposed in terms rotations around the $Y$ and $Z$ axis of the Bloch sphere.
\begin{equation}
    U(\Theta, \Phi, \lambda)= R_Z(\Phi)R_Y(\Theta)R_Z(\lambda),
\end{equation}
where $R_Y(\xi) = e^{-i \sigma_y \xi/2}$ and $R_Z(\xi) = e^{-i \sigma_z \xi/2} $, $\sigma_x, \sigma_y, ~\mbox{and}~ \sigma_z$ are the three Pauli matrices.

Physically, a single qubit gate is applied by driving the qubit through a pulse on resonance with the qubit frequency. Such a pulse produces a rotation around an axis on the equator of the Bloch sphere defined by the vector $\hat{n}= (\cos \Phi, -\sin \Phi, 0)$. The general form of such a rotation is given by
\begin{align}\label{pulse-1}
    R_{\hat{n}(\Phi)}(\Theta)= & \exp\Big(- \frac{i}{2} \Theta \hat{n}. \vec\sigma \Big)  = e^{i \frac{\Phi}{2}\sigma_z}e^{-i\frac{\Theta}{2}\sigma_y}e^{-i \frac{ \Phi}{2}\sigma_z} \nonumber \\
    = & R_Z(- \Phi) R_Y(\Theta) R_Z(\Phi) \nonumber \\
    = &  U(\Theta, - \Phi, \Phi),
\end{align}
where the vector $\vec \sigma= (\sigma_x, \sigma_y, \sigma_z)$. The parameter $\Theta$ is determined by the amplitude and duration of the pulse, while $\Phi$ corresponds to the phase of the pulse. Thus, a physical pulse implements a particular case of the generic unitary gate \eqref{ugate}.

The simulation of neutrino oscillation on a quantum circuit would require us to construct a sequence of gates that implement the desired unitary propagator of oscillation. For example, the oscillation probability \eqref{prob-1} can be simulated by implementing the unitary: $\mathscr{U}(\theta_m) \mathcal{U}(\phi) \mathscr{U}^\dagger(\theta_m)$ through an appropriate set of gates. The matrix $\mathscr{U}(\theta_m)$ corresponds to qubit rotation about the $Y$axis of the Bloch sphere and can be encoded in the quantum gate \eqref{ugate} as: $\mathscr{U}(\theta_m)\equiv U(2\theta_m, 0, 0)= R_Y(2 \theta_m)$. Physically, the $Y$ gate is executed by driving the qubit on resonance using a pulse of appropriate amplitude and duration and zero phase. The evolution matrix $\mathcal{U}(\phi)$, corresponds to qubit rotation around the $Z$ axis: $\mathcal{U}(\phi) \equiv  R_Z(\phi)$. The $Z$ gate can be applied by detuning the qubit frequency with respect to the drive pulse. Another way to implement the $Z$ gate is by rotating the qubit state using a combination of $X$ and $Y$ gates since $R_Z(\phi) = R_X(\pi/2) R_Y(\phi) R_X(- \pi/2)$. An equivalent and efficient way to do this operation is to rotate the axes of the Bloch sphere with respect to the qubit state. In practice,  this can be done by adding a phase offset to the driving pulse for all subsequent $X$ and $Y$ gates \cite{McKay:2017rej}. This phase offset does not require any physical pulse and can be done at the software level. This ``virtual" $Z$ (\texttt{VZ}) gate is free from calibration errors and has zero gate duration.

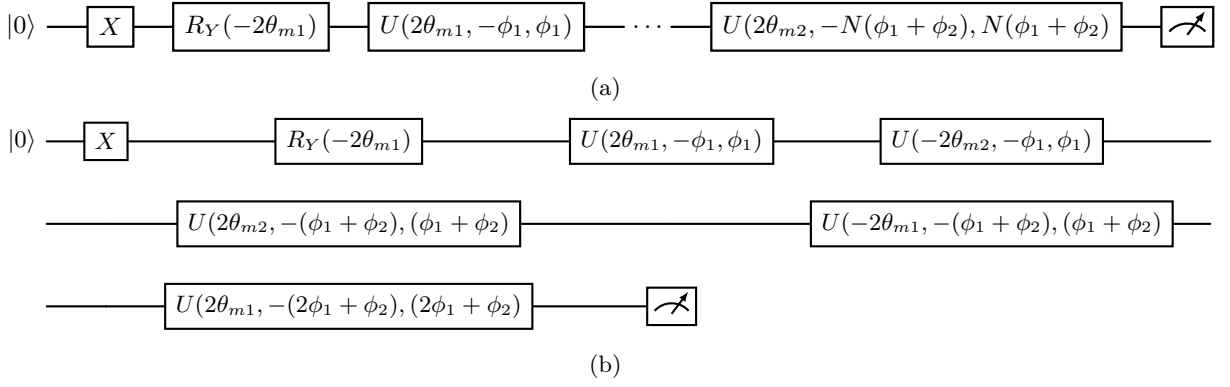
\begin{figure*}[t]
\centering
\begin{subfigure}[b]{1.0\textwidth}
\scalebox{1.06}{
\begin{quantikz}
\lstick{$\ket{0}$}&\gate{X}&\gate{R_Y(-2 \theta_{m1})}&\gate{U(2 \theta_{m1}, - \phi_1, \phi_1)}& \ \ldots\ &\gate{U(2 \theta_{m2}, - N(\phi_1+ \phi_2), N(\phi_1+ \phi_2)}& \meter{} 
\end{quantikz}}
\caption{}  \label{fig:slab-circuit}
\end{subfigure} 
\begin{subfigure}[b]{1.0\textwidth} 
\begin{quantikz}
\lstick{$\ket{0}$}&\gate{X}&\gate{R_Y(-2 \theta_{m1})}&\gate{U(2 \theta_{m1}, - \phi_1, \phi_1)}&\gate{U(-2 \theta_{m2}, -\phi_1, \phi_1)}& \\ &  &\gate{U(2 \theta_{m2}, -(\phi_1+ \phi_2), (\phi_1+ \phi_2)}&    & \gate{U(-2 \theta_{m1}, -(\phi_1+ \phi_2), (\phi_1+ \phi_2)}& \\ & \phantomgate{}&\gate{U(2 \theta_{m1}, - (2\phi_1+ \phi_2), (2\phi_1+ \phi_2)}& \meter{} 
\end{quantikz}
\caption{}\label{fig:earth-circuit}
\end{subfigure}

\caption[]{\RaggedRight (a) Implementation of Eq \eqref{slab-vz} in a quantum circuit. The $X$ gate initializes the qubit in state $\ket{1}  \equiv \ket{\nu_\mu}.$ Each unitary gate can be realized with a single physical pulse, so the circuit implementation in a QPU would require $N+1$ pulses. (b) Quantum circuit for evaluating $P(\nu_\mu \rightarrow \nu_e)$ for neutrinos passing through the center of Earth with the density profile shown in Figure \ref{fig:earth}. }
\end{figure*}

To see \texttt{VZ} gate in action, consider the following gate implementation \cite{Rajpoot:2025nzu} 
\begin{align} \label{single-slab}
    \mathscr{U}(\theta_m) & \mathcal{U}(\phi)\mathscr{U}^\dagger(\theta_m)  \equiv U(2\theta_{m}, 0,0) R_Z(\phi) U(-2\theta_{m}, 0 ,0 ) \nonumber \\ =&  R_Z(\phi) R_Z(- \phi)R_Y(2\theta_{m}) R_Z(\phi)  R_Y(-2\theta_{m}) \nonumber \\ = & R_Z( \phi)U(2\theta_{m}, - \phi,\phi)  R_Y(-2\theta_{m}),
\end{align}
where we have used Eq.\eqref{pulse-1}. The last $R_Z$ gate is irrelevant if the measurements are done in the $Z$ basis. The main advantage we obtain here by applying \texttt{VZ} gates is the reduction in the number of physical pulses required to implement the desired circuit.  This can be especially important in the case where a large number of $Z$ gates are required. For example, in Eq \eqref{slab} the case where neutrinos propagate through a sequence of $N$ slabs,  the unitary propagator requires a minimum of $3N$ physical pulses including $N$ instances of the $Z$ gate. Implementing this unitary using virtual Z gate would require only $2N$ physical pulses:
\begin{align} \label{slab-vz}
    & [\mathscr{U}(\theta_{m2}) \mathcal{U}(\phi_2) \mathscr{U}^\dagger(\theta_{m2})]_{(N)} ...[\mathscr{U}(\theta_{m2}) \mathcal{U}(\phi_2) \nonumber  \mathscr{U}(\theta_{m2})]_{(2)}\\&[\mathscr{U}(\theta_{m1}) \mathcal{U}(\phi_1) \mathscr{U}^\dagger(\theta_{m1})]_{(1)} \nonumber \\& \equiv R_Z(N(\phi_1+ \phi_2)) U(2 \theta_{m2}, -N(\phi_1+ \phi_2), N(\phi_1+ \phi_2)) \nonumber \\& U(-2 \theta_{m2}, -N\phi_1-(N-1) \phi_2, N\phi_1+(N-1) \phi_2)... \nonumber \\& U(2 \theta_{m2}, -(\phi_1+ \phi_2), (\phi_1+ \phi_2)) U(-2 \theta_{m2}, -\phi_1, \phi_1) \nonumber \\& U(2 \theta_{m1}, -\phi_1, \phi_1) R_Y(-2\theta_{m1}).
\end{align}

\begin{figure}
\centering
   \includegraphics[width=1\linewidth]{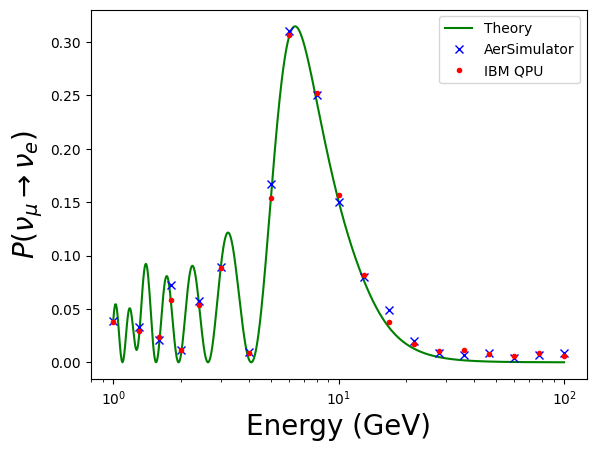}
   \caption{\RaggedRight The neutrino oscillation probability corresponding to the slab density profile shown in Figure \ref{fig:slab}. The simulations were performed using five periods ($N=10$) of representative path lengths $\Delta x_1 = 500$ km and $\Delta x_2= 1000$ km with respective densities $\rho_1= 5$ g/cm$^3$ and $\rho_2= 10$ g/cm$^3$. The effective mixing angle in each constant density slab is given by $\theta_m= \sin ^{-1}(\sin \theta_{23} \sin 2 \theta_{13m})$ \cite{Super-Kamiokande:2006jvq}, where $\theta_{23}= 45^\circ$ and $\theta_{13}= 9^\circ$. The data points obtained from Qiskit AerSimulator (blue cross) and IBM QPU (red dot) show good agreement with the theory (solid lines). }\label{fig:slab-prob}  
\end{figure}

\begin{figure}
\centering
   \includegraphics[width=1\linewidth]{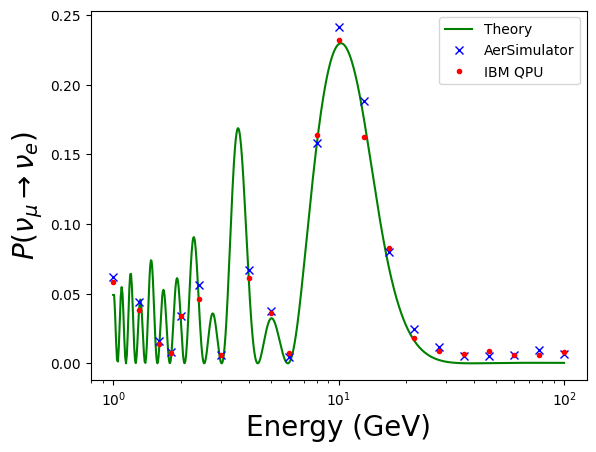}
   \caption{\RaggedRight The neutrino oscillation probability as a function of Energy for neutrinos passing through the center of the Earth. The solid line shows the theoretical calculation corresponding to the density profile shown in Figure
   \ref{fig:earth}. The quantum simulation results obtained from the Qiskit Aersimulator and IBM Quantum processor are shown with blue cross markers and red dot markers, respectively.}\label{fig:earth-prob}
\end{figure}

In Figure \ref{fig:slab-circuit}, we show the quantum circuit corresponding to Eq. \eqref{slab-vz}.  It consists of an $X$ gate, which initializes the qubit in the state $\ket{\nu_\mu}$. This is followed by a $Y$ gate and then a series of unitary operations with varying phases which effectively implement the \texttt{VZ} gates. Thus the circuit contains a total of $2N+1$ gates, each of which can be implemented by a physical pulse. The measurement at the end of the circuit provides the probabilities of the two possible outcomes, corresponding to the states $\ket{\nu_e}$ and $\ket{\nu_\mu}$, thus determining the neutrino flavor. The oscillation probability obtained by running this circuit is plotted in Figure \ref{fig:slab-prob}. The quantum circuit simulating the oscillation dynamics of atmospheric neutrinos as they propagate through the center of the Earth is shown in Figure \ref{fig:earth-circuit}. The corresponding oscillation probability is plotted in Figure \ref{fig:earth-prob}. Both circuits are executed for 4096 shots, and the oscillation probability $P(\nu_\mu \rightarrow \nu_e)$ is determined by counting the fraction of $\ket{0}$ outcomes. The results obtained from both the AerSimulator and the QPU exhibit good agreement with theoretical predictions.

Previous quantum simulation of neutrino oscillations—such as those in Refs. \cite{Arguelles:2019phs, Nguyen:2022snr, Singh:2024vpu}—were primarily restricted to vacuum oscillations or propagation through a single slab of constant-density matter. In these scenarios, implementing a physical $Z$ gate is computationally inexpensive and poses minimal overhead. For example, Ref. \cite{Arguelles:2019phs} simulated both two- and three-flavor oscillations using standard gates from the Qiskit circuit library, with time evolution encoded via the Phase gate, which is functionally equivalent to the \texttt{VZ} gate up to a global phase. In Ref. \cite{Nguyen:2022snr}, the simulation was carried out using qutrit-based gates, where the unitary evolution operator was decomposed into elementary rotations in the $\{01\}$ and $\{12\}$ subspaces. This decomposition led to a more efficient implementation by reducing the total number of gates required. In contrast, Ref. \cite{Singh:2024vpu} utilized a numerically optimized GRAPE pulse to simulate the entire unitary operator corresponding to the single-slab evolution [Eq. \eqref{single-slab}] on an NMR processor.

\begin{figure}
\centering
   \includegraphics[width=1\linewidth]{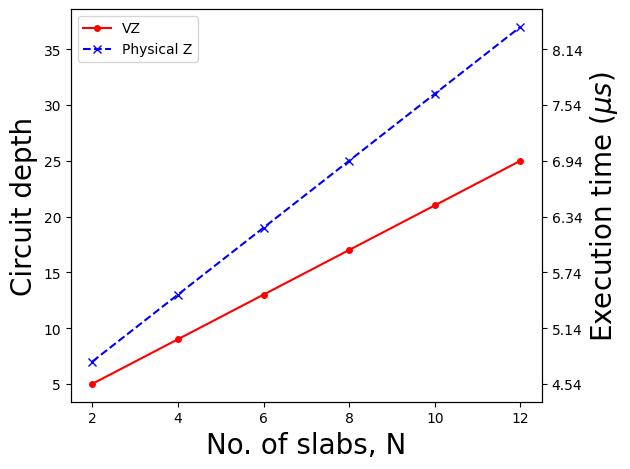}
   \caption{\RaggedRight Benchmark comparison of circuit depth and execution time for two implementations: one using the \texttt{VZ} gate and the other using physical $Z$ gates. While both circuits exhibit the same scaling behavior, the implementation with \texttt{VZ} gates has a reduced prefactor. Circuit execution time was obtained using the Qiskit \texttt{schedule} function.. }\label{fig:benchmark}  
\end{figure}

To demonstrate the novelty of our gate arrangement, we perform a benchmarking comparison between our circuit and an alternative implementation that uses physical $Z$ gates.  In contrast to earlier works, our scheme targets more general evolution scenarios where implementing multiple physical $Z$ gates becomes increasingly costly. We compare two key metrics: circuit depth and circuit execution time (see Figure \ref{fig:benchmark}). As discussed in the previous section, the circuit depth of our implementation scales as $O(2N)$, whereas for a circuit with physical $Z$ gates it scales as $O(3N)$. This indicates an improved scaling behavior with a reduced prefactor in our approach. The circuit execution time exhibits a similar trend, reflecting the efficiency of our gate arrangement in practical implementations. These improvements make the approach more amenable to near-term devices with limited coherence and gate fidelity. 


In this work, we have designed quantum circuits using gate arrangements that closely reflect the quantum operations underlying the actual process of neutrino oscillation. However, this may not be the most efficient way to implement the given unitary operation on a quantum computer. This is due to the fact that each quantum computing hardware has a distinct set of native gates and qubit connectivity constraints. In the Qiskit simulator, a given input circuit is transpiled to a form that can be executed on a specific quantum device \cite{qiskit2024}. In Appendix \ref{appex-1}, we show the transpiled form of the above circuit.

\section{\label{sec:4} Simulating solar neutrino propagation using two-qubit gates}
Solar neutrinos, generated near the Sun's core, undergo the MSW effect as they propagate outward adiabatically. Their survival probability in vacuum at a given point is determined by Eq. \eqref{prob-2}. The mixing matrices of vacuum and matter in Eq. \eqref{prob-2} have the following form
\begin{equation}
    W_{\rm vac}= \begin{pmatrix}
        \cos^2 \theta & \sin^2 \theta \\
        \sin^2 \theta & \cos^2 \theta
    \end{pmatrix},~ W_{\rm mat}=  \begin{pmatrix}
        \cos^2 \theta_m & \sin^2 \theta_m \\
        \sin^2 \theta_m & \cos^2 \theta_m
    \end{pmatrix},
\end{equation}
To simulate solar neutrino propagation, the quantum circuit must encode the matrix $Q= W_{\rm vac} W_{\rm mat}$ using a suitable arrangement of quantum gates. However, it is clear that the above matrices are not unitary, so we cannot apply quantum gates in a straightforward manner.  To encode these non-unitary matrices, we consider a composite system of two qubits consisting of an ancilla qubit ($q_A$) and an encoded qubit ($q_B$). The encoded qubit has $\ket{0} = \ket{\nu_e}$ and $\ket{{1}}= \ket{\nu_\mu}$. The time evolution of the two qubit system is engineered using the following unitary matrix: 
\begin{equation} \label{U2q}
    \mathscr{U}_{2q}= \begin{pmatrix}
        Q & \sqrt{I - Q^2} \\
        \sqrt{I- Q^2} & -Q
    \end{pmatrix},
\end{equation}
where $I$ is the $2\times2$ identity matrix. Both the qubits are initialized in the state $\ket{0}$, so the initial density matrix of the system is $\rho_0^{AB}= \ket{00} \bra{00}$. Under the unitary evolution \ref{U2q}, the density matrix of the system transforms as 
\begin{equation}
    \rho^{AB} = \mathscr{U}_{2q} \rho_0^{AB} \mathscr{U}^\dagger_{2q}.
\end{equation}
The reduced density matrix for the qubit $B$ is then given by 
\begin{equation}
    \rho^B = \mbox{tr}_A(\rho^{AB}) = \begin{pmatrix}
        Q_{00} & 0 \\ 0 & 1- Q_{00}
    \end{pmatrix},
\end{equation}
where $Q_{00} = \frac{1}{2}(1+ \cos 2 \theta \cos 2 \theta_m)$ is the element of the matrix $Q$. The measurements on the qubit $q_B$ is defined by the operators: $M_0= \ket{0} \bra{0}$ and $M_1= \ket{1} \bra{1}$. This gives us the probabilities of obtaining the outcomes $\ket{0}$ and $\ket{1}$, respectively, as: 
\begin{align}
   & p(0) = \mbox{tr}(M_0 \rho^B)= Q_{00} = P(\nu_e \rightarrow \nu_e), \nonumber \\
   &  p(1)= \mbox{tr}(M_1 \rho^B) = 1- Q_{00} = P(\nu_e \rightarrow \nu_\mu).
\end{align}
Thus we obtain the desired probability \eqref{prob-2}, by performing measurements on the encoded qubit. 

There are several ways in which the unitary $\mathscr{U}_{2q}$ implemented in a quantum circuit. This would require decomposing $\mathscr{U}_{2q}$ in a sequence of single and two qubit gates. Since $\mathscr{U}_{2q}$ is a purely real unitary matrix, it can be constructed using two \texttt{CNOT} gates \cite{Vidal:2004akd, Vatan:2004nmz}. We thus realize $\mathscr{U}_{2q}$ in a quantum circuit using two \texttt{CNOT}s and six parametrized single qubit gates \cite{Arguelles:2019phs} as follows:
\begin{align} \label{2-qubit-param}
    \mathscr{U}_{2q} \equiv &  (R_Y^A(\alpha_3)\otimes R_Y^B(\beta_3)) U_{\rm CX}^{AB}(R_Y^A(\alpha_2)\otimes R_Y^B(\beta_2)) U_{\rm CX}^{AB} \nonumber \\&(R_Y^A(\alpha_1)\otimes R_Y^B(\beta_1)),
\end{align}
where $U_{CX}^{AB}$ is the \texttt{CNOT} gate with control qubit $q_A$ and target qubit $q_B$. The parameters $\alpha_i, \beta_i~ (i= 1,2, 3)$ are obtained by optimizing the gate fidelity (Appendix \ref{appex-2}). 

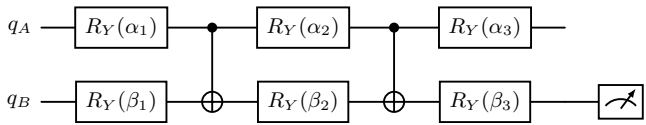
\begin{figure}
\scalebox{0.9}{
\begin{quantikz}
\lstick{$q_A$}&\gate{R_Y(\alpha_1)}& \ctrl{1} & \gate{R_Y(\alpha_2)}& \ctrl{1}& \gate{R_Y(\alpha_3)} & \\ \lstick{$q_B$}&\gate{R_Y(\beta_1)} & \targ{2}&\gate{R_Y(\beta_2)}& \targ{2} &\gate{R_Y(\beta_3)} & & \meter{}
\end{quantikz}}
\caption{\RaggedRight Quantum circuit implementing Eq. \eqref{2-qubit-param}. The qubit $q_B$ encodes the neutrino flavor states, and $q_A$ is the ancilla qubit. The parameters $\alpha_i$ and $\beta_i$ $(i= 1,2,3)$ are optimized for high gate fidelity. The measurements are performed on the qubit $q_B$.} \label{2-qubit-circuit}
\end{figure}

\begin{figure}
\centering
   \includegraphics[width=1\linewidth]{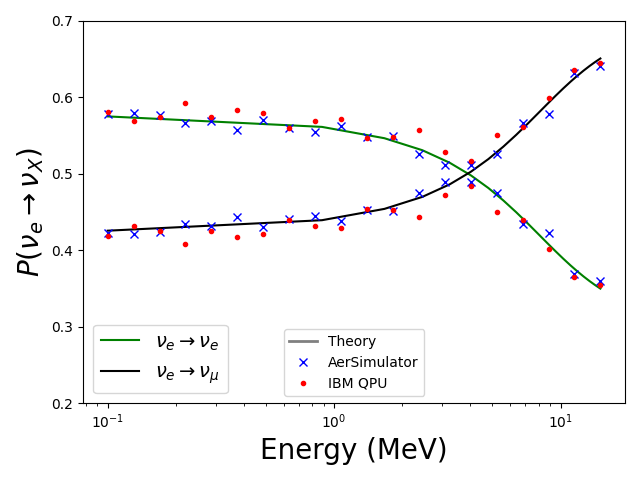}
   \caption{\RaggedRight Survival $\nu_e \rightarrow \nu_e$ and transition $\nu_e \rightarrow\nu_\mu$ probabilities of $\nu_e$ as a function of energy for solar neutrinos. The solid lines represent the theoretical curve corresponding to Eq. \eqref{prob-2}, with $\theta= \theta_{12}= 33.5^\circ$. The blue cross markers and red dot markers correspond to data obtained from the Qiskit AerSimulator and the IBM QPU, respectively.}\label{fig:msw-prob}
\end{figure}

The quantum circuit corresponding to Eq. \eqref{2-qubit-param}  is shown in Figure \ref{2-qubit-circuit}. The two qubits are initialized in the state $\ket{0}$ and a sequence of single and two qubit gates is applied to encode the desired evolution. Finally, the measurements are made on the qubit $q_B$. The corresponding results obtained after the measurement are plotted in Figure \ref{fig:msw-prob}. For each data point, we execute the circuit 4096 times to buid sufficient statistics. Both the  measurement results $P(\nu_e \rightarrow \nu_e)$ and $P(\nu_e \rightarrow \nu_\mu)$ show good agreement with the theoretical prediction.    

\section{\label{sec:5}Conclusions}
We have performed quantum simulations of neutrino propagation in both constant and varying matter density profile. The quantum circuits for simulating two flavor neutrino oscillations through periodic density slab and through Earth matter was presented. These circuits employ a sequence of single qubit gates along with \texttt{VZ} gates.Benchmarking comparison show that the  use of \texttt{VZ} gates reduces the overall circuit depth and execution time, which is important for executing algorithms on noisy qubits. The propagation of neutrinos through a varying density profile was simulated for the case of solar neutrinos. Here, due to the non-unitary nature of the evolution matrix, the quantum circuit requires two qubit entangling gates to simulate the desired probabilities. For all the circuits discussed above, the simulations were performed on both the Qiskit AerSimulator and the IBM QPU hardware. The results obtained from quantum simulations show a good agreement with the theoretical calculations.      

The quantum simulation of neutrino propagation presented here shows the versatility of the present day quantum simulators and QPUs. In principle, any complex quantum system to be simulated can be mapped into qubit states and the evolution of this simulated system can be encoded in terms of an effective unitary propagator. This unitary can be implemented on a quantum circuit to an arbitrary accuracy using a discrete set of single and two qubit gates \cite{Nielsen:2012yss}. In the case of neutrinos, this strategy can be utilized to simulate various other propagation scenarios involving spin-flavor oscillations \cite{Giunti:2014ixa, Joshi:2017vpi}, quantum decoherence \cite{Benatti:2000ph, Oliveira:2010zzd}, non-standard interactions \cite{Ohlsson:2012kf}, etc.

\section*{Acknowledgements}
We acknowledge the use of Qiskit AerSimulator and \texttt{ibm\_brisbane} in this study. The simulations were perfromed using Qiskit version 1.3.0 and Qiskit-Aer version 0.15.1. The views expressed are those of the authors and do not reflect the official policy or position of IBM or the IBM Quantum team. 
\appendix

\section{Qiskit transpilation of quantum circuits} \label{appex-1}
Qiskit transpiler rewrites a given input circuit in order to match the instruction set architecture (ISA) of the target backend. The transformed circuit obeying the ISA consists only of instructions supported by the backend device, such as the hardware's available basis gates, measurements, resets, and control flow operations, and comply with the constraints specified by the connectivity of the hardware. In addition, the transpiler optimizes the circuit instructions, which greatly reduces the depth and complexity of quantum circuits. In this paper, we used \texttt{ibm\_brisbane}, which is one of the IBM Eagle processors. The native gate set available in this processor include \texttt{ECR} (echoed cross-resonance, two qubit gate), \texttt{RZ} (Single qubit $Z$ rotation), $\sqrt{\texttt{X}}$ (single qubit $\sqrt{\texttt{NOT}}$ gate), \texttt{X} (single qubit \texttt{NOT} gate) and \texttt{ID} (single qubit \texttt{Identity}) gates.

\begin{figure}
\centering
   \includegraphics[width= \linewidth]{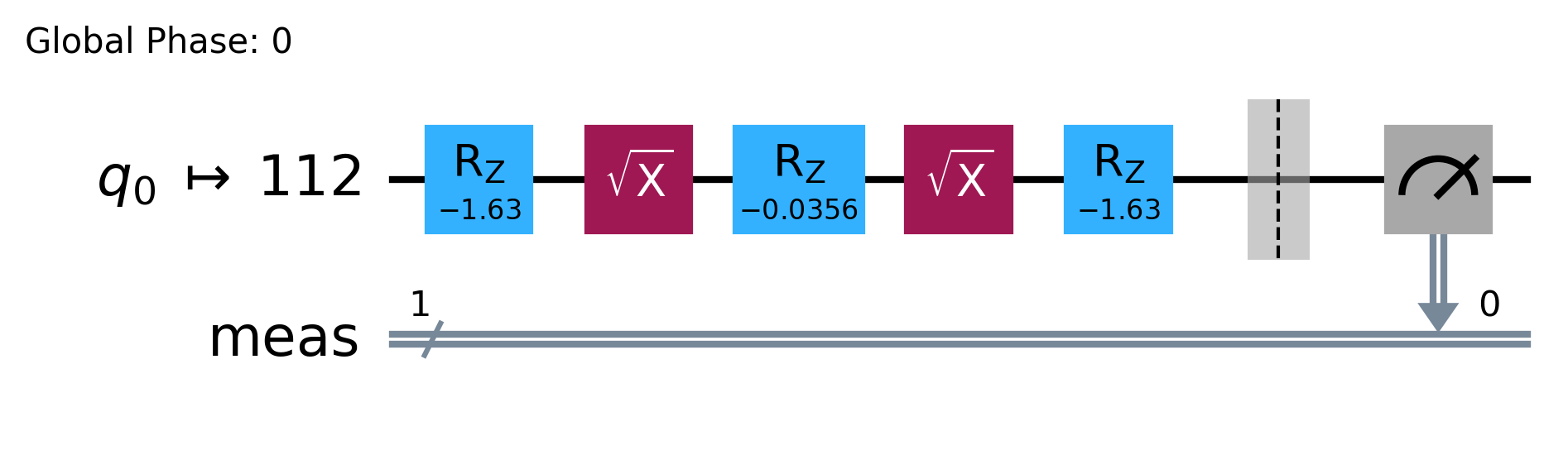}
   \caption{\RaggedRight Quantum circuit after transpilation corresponding to the input circuit shown in Figure \ref{fig:earth-circuit}. }\label{fig:isa-earth}
   \label{fig:msw} 
\end{figure}

In Figure \ref{fig:isa-earth}, we show the transpiled circuit corresponding to the circuit shown in Figure \ref{fig:earth-circuit}, which simulates the neutrino propagation through Earth. The transpiler has clearly reduced the gate count, which plays an important role in getting better results from the noisy qubits. The transpiler has an option \texttt{optimization\_level}, which takes integer values 0, 1, 2, or 3. Choosing a higher optimization level generates more optimized circuits, however, it takes longer to compile such circuits. In this paper, we set the \texttt{optimization\_level} equal to 2 for all the circuits. 

\section{Optimal parameters for single qubit gates} \label{appex-2}
The parameters $\alpha_i$ and $\beta_i$ in Eq. \eqref{2-qubit-param} are obtained by maximizing the overlap fidelity
\begin{equation}
    F = \frac{1}{16}|\mbox{Tr}(U^\dagger_T U_R )|^2, 
\end{equation}
where $U_T$ is the target unitary given by Eq. \eqref{U2q} and $U_R$ is the realized unitary \eqref{2-qubit-param}. We optimize the parameters using the \texttt{optimize} module from \texttt{Scipy}. The parameters $\alpha_i$ and $\beta_i$ are constrained within the range $(-\pi, \pi)$, and the optimization is performed using the \texttt{L-BFGS-B} method. Since this is a gradient-based optimization algorithm, it is prone to falling into local minima. To mitigate this, we randomize the initial parameter guesses within the range $(-1, 1)$ and repeat the optimization 1000 times for each data point. We obtain an infidelity $1- F$  less than $10^{-7}$ for all the data points. For example for the first data point in Figure \ref{fig:msw-prob}, we obtain $1-F = 4.5 \times 10^{-9}$ and the optimized parameters are: $\alpha_1  = 0.89140528, ~ \beta_1= 1.57079633,~ \alpha_2=  1.42089489 ,~ \beta_2=   -3.14159265,~ \alpha_3=  -0.82929248,~ \beta_3 = 1.57079633$.  

\bibliography{zzbooks,zznuosc, zznuexpt,zzsolar,zzquantum}

\begin{thebibliography}{56}%
\makeatletter
\providecommand \@ifxundefined [1]{%
 \@ifx{#1\undefined}
}%
\providecommand \@ifnum [1]{%
 \ifnum #1\expandafter \@firstoftwo
 \else \expandafter \@secondoftwo
 \fi
}%
\providecommand \@ifx [1]{%
 \ifx #1\expandafter \@firstoftwo
 \else \expandafter \@secondoftwo
 \fi
}%
\providecommand \natexlab [1]{#1}%
\providecommand \enquote  [1]{``#1''}%
\providecommand \bibnamefont  [1]{#1}%
\providecommand \bibfnamefont [1]{#1}%
\providecommand \citenamefont [1]{#1}%
\providecommand \href@noop [0]{\@secondoftwo}%
\providecommand \href [0]{\begingroup \@sanitize@url \@href}%
\providecommand \@href[1]{\@@startlink{#1}\@@href}%
\providecommand \@@href[1]{\endgroup#1\@@endlink}%
\providecommand \@sanitize@url [0]{\catcode `\\12\catcode `\$12\catcode `\&12\catcode `\#12\catcode `\^12\catcode `\_12\catcode `\%12\relax}%
\providecommand \@@startlink[1]{}%
\providecommand \@@endlink[0]{}%
\providecommand \url  [0]{\begingroup\@sanitize@url \@url }%
\providecommand \@url [1]{\endgroup\@href {#1}{\urlprefix }}%
\providecommand \urlprefix  [0]{URL }%
\providecommand \Eprint [0]{\href }%
\providecommand \doibase [0]{https://doi.org/}%
\providecommand \selectlanguage [0]{\@gobble}%
\providecommand \bibinfo  [0]{\@secondoftwo}%
\providecommand \bibfield  [0]{\@secondoftwo}%
\providecommand \translation [1]{[#1]}%
\providecommand \BibitemOpen [0]{}%
\providecommand \bibitemStop [0]{}%
\providecommand \bibitemNoStop [0]{.\EOS\space}%
\providecommand \EOS [0]{\spacefactor3000\relax}%
\providecommand \BibitemShut  [1]{\csname bibitem#1\endcsname}%
\let\auto@bib@innerbib\@empty
\bibitem [{\citenamefont {Fukuda}\ \emph {et~al.}(1998)\citenamefont {Fukuda} \emph {et~al.}}]{Fukuda:1998mi}%
  \BibitemOpen
  \bibfield  {author} {\bibinfo {author} {\bibfnamefont {Y.}~\bibnamefont {Fukuda}} \emph {et~al.} (\bibinfo {collaboration} {Super-Kamiokande}),\ }\bibfield  {title} {\bibinfo {title} {{Evidence for oscillation of atmospheric neutrinos}},\ }\href {https://doi.org/10.1103/PhysRevLett.81.1562} {\bibfield  {journal} {\bibinfo  {journal} {Phys. Rev. Lett.}\ }\textbf {\bibinfo {volume} {81}},\ \bibinfo {pages} {1562} (\bibinfo {year} {1998})},\ \Eprint {https://arxiv.org/abs/hep-ex/9807003} {arXiv:hep-ex/9807003 [hep-ex]} \BibitemShut {NoStop}%
\bibitem [{\citenamefont {Ahmad}\ \emph {et~al.}(2002)\citenamefont {Ahmad} \emph {et~al.}}]{Ahmad:2002jz}%
  \BibitemOpen
  \bibfield  {author} {\bibinfo {author} {\bibfnamefont {Q.~R.}\ \bibnamefont {Ahmad}} \emph {et~al.} (\bibinfo {collaboration} {SNO}),\ }\bibfield  {title} {\bibinfo {title} {{Direct evidence for neutrino flavor transformation from neutral current interactions in the Sudbury Neutrino Observatory}},\ }\href {https://doi.org/10.1103/PhysRevLett.89.011301} {\bibfield  {journal} {\bibinfo  {journal} {Phys. Rev. Lett.}\ }\textbf {\bibinfo {volume} {89}},\ \bibinfo {pages} {011301} (\bibinfo {year} {2002})},\ \Eprint {https://arxiv.org/abs/nucl-ex/0204008} {arXiv:nucl-ex/0204008 [nucl-ex]} \BibitemShut {NoStop}%
\bibitem [{\citenamefont {Araki}\ \emph {et~al.}(2005)\citenamefont {Araki} \emph {et~al.}}]{Araki:2004mb}%
  \BibitemOpen
  \bibfield  {author} {\bibinfo {author} {\bibfnamefont {T.}~\bibnamefont {Araki}} \emph {et~al.} (\bibinfo {collaboration} {KamLAND}),\ }\bibfield  {title} {\bibinfo {title} {{Measurement of neutrino oscillation with KamLAND: Evidence of spectral distortion}},\ }\href {https://doi.org/10.1103/PhysRevLett.94.081801} {\bibfield  {journal} {\bibinfo  {journal} {Phys. Rev. Lett.}\ }\textbf {\bibinfo {volume} {94}},\ \bibinfo {pages} {081801} (\bibinfo {year} {2005})},\ \Eprint {https://arxiv.org/abs/hep-ex/0406035} {arXiv:hep-ex/0406035 [hep-ex]} \BibitemShut {NoStop}%
\bibitem [{\citenamefont {Giunti}\ and\ \citenamefont {Kim}(2007)}]{Giunti:2007ry}%
  \BibitemOpen
  \bibfield  {author} {\bibinfo {author} {\bibfnamefont {C.}~\bibnamefont {Giunti}}\ and\ \bibinfo {author} {\bibfnamefont {C.~W.}\ \bibnamefont {Kim}},\ }\href@noop {} {\emph {\bibinfo {title} {{Fundamentals of Neutrino Physics and Astrophysics}}}}\ (\bibinfo  {publisher} {Oxford University Press},\ \bibinfo {year} {2007})\BibitemShut {NoStop}%
\bibitem [{\citenamefont {Joshi}\ and\ \citenamefont {Jain}(2020)}]{Joshi:2019dcj}%
  \BibitemOpen
  \bibfield  {author} {\bibinfo {author} {\bibfnamefont {S.}~\bibnamefont {Joshi}}\ and\ \bibinfo {author} {\bibfnamefont {S.~R.}\ \bibnamefont {Jain}},\ }\bibfield  {title} {\bibinfo {title} {{Neutrino spin-flavor oscillations in solar environment}},\ }\href {https://doi.org/10.1088/1674-4527/20/8/123} {\bibfield  {journal} {\bibinfo  {journal} {Res. Astron. Astrophys.}\ }\textbf {\bibinfo {volume} {20}},\ \bibinfo {pages} {123} (\bibinfo {year} {2020})},\ \Eprint {https://arxiv.org/abs/1906.09475} {arXiv:1906.09475 [hep-ph]} \BibitemShut {NoStop}%
\bibitem [{\citenamefont {Kuo}\ and\ \citenamefont {Pantaleone}(1989)}]{Kuo:1989qe}%
  \BibitemOpen
  \bibfield  {author} {\bibinfo {author} {\bibfnamefont {T.-K.}\ \bibnamefont {Kuo}}\ and\ \bibinfo {author} {\bibfnamefont {J.~T.}\ \bibnamefont {Pantaleone}},\ }\bibfield  {title} {\bibinfo {title} {{Neutrino Oscillations in Matter}},\ }\href {https://doi.org/10.1103/RevModPhys.61.937} {\bibfield  {journal} {\bibinfo  {journal} {Rev. Mod. Phys.}\ }\textbf {\bibinfo {volume} {61}},\ \bibinfo {pages} {937} (\bibinfo {year} {1989})}\BibitemShut {NoStop}%
\bibitem [{\citenamefont {Joshi}(2020)}]{Joshi:2020djx}%
  \BibitemOpen
  \bibfield  {author} {\bibinfo {author} {\bibfnamefont {S.}~\bibnamefont {Joshi}},\ }\bibfield  {title} {\bibinfo {title} {{Mixed state geometric phase for neutrino oscillations}},\ }\href {https://doi.org/10.1016/j.physletb.2020.135766} {\bibfield  {journal} {\bibinfo  {journal} {Phys. Lett. B}\ }\textbf {\bibinfo {volume} {809}},\ \bibinfo {pages} {135766} (\bibinfo {year} {2020})},\ \Eprint {https://arxiv.org/abs/2008.07952} {arXiv:2008.07952 [hep-ph]} \BibitemShut {NoStop}%
\bibitem [{\citenamefont {Feynman}(1982)}]{Feynman:1981tf}%
  \BibitemOpen
  \bibfield  {author} {\bibinfo {author} {\bibfnamefont {R.~P.}\ \bibnamefont {Feynman}},\ }\bibfield  {title} {\bibinfo {title} {{Simulating physics with computers}},\ }\href {https://doi.org/10.1007/BF02650179} {\bibfield  {journal} {\bibinfo  {journal} {Int. J. Theor. Phys.}\ }\textbf {\bibinfo {volume} {21}},\ \bibinfo {pages} {467} (\bibinfo {year} {1982})}\BibitemShut {NoStop}%
\bibitem [{\citenamefont {Lloyd}(1996)}]{Lloyd:1996aai}%
  \BibitemOpen
  \bibfield  {author} {\bibinfo {author} {\bibfnamefont {S.}~\bibnamefont {Lloyd}},\ }\bibfield  {title} {\bibinfo {title} {{Universal Quantum Simulators}},\ }\href {https://doi.org/10.1126/science.273.5278.1073} {\bibfield  {journal} {\bibinfo  {journal} {Science}\ }\textbf {\bibinfo {volume} {273}},\ \bibinfo {pages} {1073} (\bibinfo {year} {1996})}\BibitemShut {NoStop}%
\bibitem [{\citenamefont {Georgescu}\ \emph {et~al.}(2014)\citenamefont {Georgescu}, \citenamefont {Ashhab},\ and\ \citenamefont {Nori}}]{Georgescu:2013oza}%
  \BibitemOpen
  \bibfield  {author} {\bibinfo {author} {\bibfnamefont {I.~M.}\ \bibnamefont {Georgescu}}, \bibinfo {author} {\bibfnamefont {S.}~\bibnamefont {Ashhab}},\ and\ \bibinfo {author} {\bibfnamefont {F.}~\bibnamefont {Nori}},\ }\bibfield  {title} {\bibinfo {title} {{Quantum Simulation}},\ }\href {https://doi.org/10.1103/RevModPhys.86.153} {\bibfield  {journal} {\bibinfo  {journal} {Rev. Mod. Phys.}\ }\textbf {\bibinfo {volume} {86}},\ \bibinfo {pages} {153} (\bibinfo {year} {2014})},\ \Eprint {https://arxiv.org/abs/1308.6253} {arXiv:1308.6253 [quant-ph]} \BibitemShut {NoStop}%
\bibitem [{\citenamefont {Johnson}\ \emph {et~al.}(2014)\citenamefont {Johnson}, \citenamefont {Clark},\ and\ \citenamefont {Jaksch}}]{Johnson:2014igy}%
  \BibitemOpen
  \bibfield  {author} {\bibinfo {author} {\bibfnamefont {T.~H.}\ \bibnamefont {Johnson}}, \bibinfo {author} {\bibfnamefont {S.~R.}\ \bibnamefont {Clark}},\ and\ \bibinfo {author} {\bibfnamefont {D.}~\bibnamefont {Jaksch}},\ }\bibfield  {title} {\bibinfo {title} {{What is a quantum simulator?}},\ }\href {https://doi.org/10.1140/epjqt10} {\bibfield  {journal} {\bibinfo  {journal} {EPJ Quant. Technol.}\ }\textbf {\bibinfo {volume} {1}},\ \bibinfo {pages} {10} (\bibinfo {year} {2014})}\BibitemShut {NoStop}%
\bibitem [{\citenamefont {Garc\'\i{}a-Ramos}\ \emph {et~al.}(2023)\citenamefont {Garc\'\i{}a-Ramos}, \citenamefont {S\'aiz}, \citenamefont {Arias}, \citenamefont {Lamata},\ and\ \citenamefont {P\'erez-Fern\'andez}}]{Garcia-Ramos:2023rtd}%
  \BibitemOpen
  \bibfield  {author} {\bibinfo {author} {\bibfnamefont {J.~E.}\ \bibnamefont {Garc\'\i{}a-Ramos}}, \bibinfo {author} {\bibfnamefont {A.}~\bibnamefont {S\'aiz}}, \bibinfo {author} {\bibfnamefont {J.~M.}\ \bibnamefont {Arias}}, \bibinfo {author} {\bibfnamefont {L.}~\bibnamefont {Lamata}},\ and\ \bibinfo {author} {\bibfnamefont {P.}~\bibnamefont {P\'erez-Fern\'andez}},\ }\bibfield  {title} {\bibinfo {title} {{Nuclear Physics in the Era of Quantum Computing and Quantum Machine Learning}},\ }\href@noop {} {\  (\bibinfo {year} {2023})},\ \Eprint {https://arxiv.org/abs/2307.07332} {arXiv:2307.07332 [quant-ph]} \BibitemShut {NoStop}%
\bibitem [{\citenamefont {Roggero}\ \emph {et~al.}(2020)\citenamefont {Roggero}, \citenamefont {Li}, \citenamefont {Carlson}, \citenamefont {Gupta},\ and\ \citenamefont {Perdue}}]{Roggero:2019myu}%
  \BibitemOpen
  \bibfield  {author} {\bibinfo {author} {\bibfnamefont {A.}~\bibnamefont {Roggero}}, \bibinfo {author} {\bibfnamefont {A.~C.~Y.}\ \bibnamefont {Li}}, \bibinfo {author} {\bibfnamefont {J.}~\bibnamefont {Carlson}}, \bibinfo {author} {\bibfnamefont {R.}~\bibnamefont {Gupta}},\ and\ \bibinfo {author} {\bibfnamefont {G.~N.}\ \bibnamefont {Perdue}},\ }\bibfield  {title} {\bibinfo {title} {{Quantum Computing for Neutrino-Nucleus Scattering}},\ }\href {https://doi.org/10.1103/PhysRevD.101.074038} {\bibfield  {journal} {\bibinfo  {journal} {Phys. Rev. D}\ }\textbf {\bibinfo {volume} {101}},\ \bibinfo {pages} {074038} (\bibinfo {year} {2020})},\ \Eprint {https://arxiv.org/abs/1911.06368} {arXiv:1911.06368 [quant-ph]} \BibitemShut {NoStop}%
\bibitem [{\citenamefont {Singh}\ \emph {et~al.}(2025)\citenamefont {Singh}, \citenamefont {Abhishek},\ and\ \citenamefont {Arumugam}}]{singh2025quantum}%
  \BibitemOpen
  \bibfield  {author} {\bibinfo {author} {\bibfnamefont {N.}~\bibnamefont {Singh}}, \bibinfo {author} {\bibnamefont {Abhishek}},\ and\ \bibinfo {author} {\bibfnamefont {P.}~\bibnamefont {Arumugam}},\ }\bibfield  {title} {\bibinfo {title} {A quantum algorithm for the linear response of nuclei},\ }\href {https://doi.org/https://doi.org/10.1007/s12648-025-03627-8} {\bibfield  {journal} {\bibinfo  {journal} {Indian Journal of Physics}\ ,\ \bibinfo {pages} {1}} (\bibinfo {year} {2025})}\BibitemShut {NoStop}%
\bibitem [{\citenamefont {Chen}\ \emph {et~al.}(2021)\citenamefont {Chen}, \citenamefont {Ma},\ and\ \citenamefont {Zhou}}]{Chen:2021cyw}%
  \BibitemOpen
  \bibfield  {author} {\bibinfo {author} {\bibfnamefont {Y.}~\bibnamefont {Chen}}, \bibinfo {author} {\bibfnamefont {Y.}~\bibnamefont {Ma}},\ and\ \bibinfo {author} {\bibfnamefont {S.}~\bibnamefont {Zhou}},\ }\bibfield  {title} {\bibinfo {title} {{Quantum Simulations of the Non-Unitary Time Evolution and Applications to Neutral-Kaon Oscillations}},\ }\href@noop {} {\  (\bibinfo {year} {2021})},\ \Eprint {https://arxiv.org/abs/2105.04765} {arXiv:2105.04765 [hep-ph]} \BibitemShut {NoStop}%
\bibitem [{\citenamefont {Bauer}\ \emph {et~al.}(2023{\natexlab{a}})\citenamefont {Bauer} \emph {et~al.}}]{Bauer:2022hpo}%
  \BibitemOpen
  \bibfield  {author} {\bibinfo {author} {\bibfnamefont {C.~W.}\ \bibnamefont {Bauer}} \emph {et~al.},\ }\bibfield  {title} {\bibinfo {title} {{Quantum Simulation for High-Energy Physics}},\ }\href {https://doi.org/10.1103/PRXQuantum.4.027001} {\bibfield  {journal} {\bibinfo  {journal} {PRX Quantum}\ }\textbf {\bibinfo {volume} {4}},\ \bibinfo {pages} {027001} (\bibinfo {year} {2023}{\natexlab{a}})},\ \Eprint {https://arxiv.org/abs/2204.03381} {arXiv:2204.03381 [quant-ph]} \BibitemShut {NoStop}%
\bibitem [{\citenamefont {Di~Meglio}\ \emph {et~al.}(2024)\citenamefont {Di~Meglio} \emph {et~al.}}]{DiMeglio:2023nsa}%
  \BibitemOpen
  \bibfield  {author} {\bibinfo {author} {\bibfnamefont {A.}~\bibnamefont {Di~Meglio}} \emph {et~al.},\ }\bibfield  {title} {\bibinfo {title} {{Quantum Computing for High-Energy Physics: State of the Art and Challenges}},\ }\href {https://doi.org/10.1103/PRXQuantum.5.037001} {\bibfield  {journal} {\bibinfo  {journal} {PRX Quantum}\ }\textbf {\bibinfo {volume} {5}},\ \bibinfo {pages} {037001} (\bibinfo {year} {2024})},\ \Eprint {https://arxiv.org/abs/2307.03236} {arXiv:2307.03236 [quant-ph]} \BibitemShut {NoStop}%
\bibitem [{\citenamefont {Bauer}\ \emph {et~al.}(2023{\natexlab{b}})\citenamefont {Bauer}, \citenamefont {Davoudi}, \citenamefont {Klco},\ and\ \citenamefont {Savage}}]{Bauer:2023qgm}%
  \BibitemOpen
  \bibfield  {author} {\bibinfo {author} {\bibfnamefont {C.~W.}\ \bibnamefont {Bauer}}, \bibinfo {author} {\bibfnamefont {Z.}~\bibnamefont {Davoudi}}, \bibinfo {author} {\bibfnamefont {N.}~\bibnamefont {Klco}},\ and\ \bibinfo {author} {\bibfnamefont {M.~J.}\ \bibnamefont {Savage}},\ }\bibfield  {title} {\bibinfo {title} {{Quantum simulation of fundamental particles and forces}},\ }\href {https://doi.org/10.1038/s42254-023-00599-8} {\bibfield  {journal} {\bibinfo  {journal} {Nature Rev. Phys.}\ }\textbf {\bibinfo {volume} {5}},\ \bibinfo {pages} {420} (\bibinfo {year} {2023}{\natexlab{b}})},\ \Eprint {https://arxiv.org/abs/2404.06298} {arXiv:2404.06298 [hep-ph]} \BibitemShut {NoStop}%
\bibitem [{\citenamefont {Noh}\ \emph {et~al.}(2012)\citenamefont {Noh}, \citenamefont {Rodr{\'\i}guez-Lara},\ and\ \citenamefont {Angelakis}}]{noh2012quantum}%
  \BibitemOpen
  \bibfield  {author} {\bibinfo {author} {\bibfnamefont {C.}~\bibnamefont {Noh}}, \bibinfo {author} {\bibfnamefont {B.}~\bibnamefont {Rodr{\'\i}guez-Lara}},\ and\ \bibinfo {author} {\bibfnamefont {D.}~\bibnamefont {Angelakis}},\ }\bibfield  {title} {\bibinfo {title} {Quantum simulation of neutrino oscillations with trapped ions},\ }\href@noop {} {\bibfield  {journal} {\bibinfo  {journal} {New Journal of Physics}\ }\textbf {\bibinfo {volume} {14}},\ \bibinfo {pages} {033028} (\bibinfo {year} {2012})}\BibitemShut {NoStop}%
\bibitem [{\citenamefont {Arg\"uelles}\ and\ \citenamefont {Jones}(2019)}]{Arguelles:2019phs}%
  \BibitemOpen
  \bibfield  {author} {\bibinfo {author} {\bibfnamefont {C.~A.}\ \bibnamefont {Arg\"uelles}}\ and\ \bibinfo {author} {\bibfnamefont {B.~J.~P.}\ \bibnamefont {Jones}},\ }\bibfield  {title} {\bibinfo {title} {{Neutrino Oscillations in a Quantum Processor}},\ }\href {https://doi.org/10.1103/PhysRevResearch.1.033176} {\bibfield  {journal} {\bibinfo  {journal} {Phys. Rev. Research.}\ }\textbf {\bibinfo {volume} {1}},\ \bibinfo {pages} {033176} (\bibinfo {year} {2019})},\ \Eprint {https://arxiv.org/abs/1904.10559} {arXiv:1904.10559 [quant-ph]} \BibitemShut {NoStop}%
\bibitem [{\citenamefont {Jha}\ and\ \citenamefont {Chatla}(2022)}]{Jha:2021itm}%
  \BibitemOpen
  \bibfield  {author} {\bibinfo {author} {\bibfnamefont {A.~K.}\ \bibnamefont {Jha}}\ and\ \bibinfo {author} {\bibfnamefont {A.}~\bibnamefont {Chatla}},\ }\bibfield  {title} {\bibinfo {title} {{Quantum studies of neutrinos on IBMQ processors}},\ }\href {https://doi.org/10.1140/epjs/s11734-021-00358-9} {\bibfield  {journal} {\bibinfo  {journal} {Eur. Phys. J. ST}\ }\textbf {\bibinfo {volume} {231}},\ \bibinfo {pages} {141} (\bibinfo {year} {2022})}\BibitemShut {NoStop}%
\bibitem [{\citenamefont {Nguyen}\ \emph {et~al.}(2023)\citenamefont {Nguyen}, \citenamefont {Bach}, \citenamefont {Nguyen}, \citenamefont {Tran}, \citenamefont {Nguyen},\ and\ \citenamefont {Nguyen}}]{Nguyen:2022snr}%
  \BibitemOpen
  \bibfield  {author} {\bibinfo {author} {\bibfnamefont {H.~C.}\ \bibnamefont {Nguyen}}, \bibinfo {author} {\bibfnamefont {B.~G.}\ \bibnamefont {Bach}}, \bibinfo {author} {\bibfnamefont {T.~D.}\ \bibnamefont {Nguyen}}, \bibinfo {author} {\bibfnamefont {D.~M.}\ \bibnamefont {Tran}}, \bibinfo {author} {\bibfnamefont {D.~V.}\ \bibnamefont {Nguyen}},\ and\ \bibinfo {author} {\bibfnamefont {H.~Q.}\ \bibnamefont {Nguyen}},\ }\bibfield  {title} {\bibinfo {title} {{Simulating neutrino oscillations on a superconducting qutrit}},\ }\href {https://doi.org/10.1103/PhysRevD.108.023013} {\bibfield  {journal} {\bibinfo  {journal} {Phys. Rev. D}\ }\textbf {\bibinfo {volume} {108}},\ \bibinfo {pages} {023013} (\bibinfo {year} {2023})},\ \Eprint {https://arxiv.org/abs/2212.14170} {arXiv:2212.14170 [quant-ph]} \BibitemShut {NoStop}%
\bibitem [{\citenamefont {Singh}\ \emph {et~al.}(2024)\citenamefont {Singh}, \citenamefont {Arvind},\ and\ \citenamefont {Dorai}}]{Singh:2024vpu}%
  \BibitemOpen
  \bibfield  {author} {\bibinfo {author} {\bibfnamefont {G.}~\bibnamefont {Singh}}, \bibinfo {author} {\bibnamefont {Arvind}},\ and\ \bibinfo {author} {\bibfnamefont {K.}~\bibnamefont {Dorai}},\ }\bibfield  {title} {\bibinfo {title} {{Simulating Three-Flavor Neutrino Oscillations on an NMR Quantum Processor}},\ }\href@noop {} {\  (\bibinfo {year} {2024})},\ \Eprint {https://arxiv.org/abs/2412.15617} {arXiv:2412.15617 [quant-ph]} \BibitemShut {NoStop}%
\bibitem [{\citenamefont {Rajpoot}\ \emph {et~al.}(2025)\citenamefont {Rajpoot}, \citenamefont {Joshi},\ and\ \citenamefont {Shukla}}]{Rajpoot:2025nzu}%
  \BibitemOpen
  \bibfield  {author} {\bibinfo {author} {\bibfnamefont {G.}~\bibnamefont {Rajpoot}}, \bibinfo {author} {\bibfnamefont {S.}~\bibnamefont {Joshi}},\ and\ \bibinfo {author} {\bibfnamefont {P.}~\bibnamefont {Shukla}},\ }\bibfield  {title} {\bibinfo {title} {{Simulating neutrino oscillations on a quantum circuit using virtual-Z gates}},\ }\href@noop {} {\bibfield  {journal} {\bibinfo  {journal} {DAE Symp. Nucl. Phys.}\ }\textbf {\bibinfo {volume} {68}},\ \bibinfo {pages} {1051} (\bibinfo {year} {2025})}\BibitemShut {NoStop}%
\bibitem [{\citenamefont {Koch}\ \emph {et~al.}(2007)\citenamefont {Koch}, \citenamefont {Yu}, \citenamefont {Gambetta}, \citenamefont {Houck}, \citenamefont {Schuster}, \citenamefont {Majer}, \citenamefont {Blais}, \citenamefont {Devoret}, \citenamefont {Girvin},\ and\ \citenamefont {Schoelkopf}}]{Koch:2007hay}%
  \BibitemOpen
  \bibfield  {author} {\bibinfo {author} {\bibfnamefont {J.}~\bibnamefont {Koch}}, \bibinfo {author} {\bibfnamefont {T.~M.}\ \bibnamefont {Yu}}, \bibinfo {author} {\bibfnamefont {J.}~\bibnamefont {Gambetta}}, \bibinfo {author} {\bibfnamefont {A.~A.}\ \bibnamefont {Houck}}, \bibinfo {author} {\bibfnamefont {D.~I.}\ \bibnamefont {Schuster}}, \bibinfo {author} {\bibfnamefont {J.}~\bibnamefont {Majer}}, \bibinfo {author} {\bibfnamefont {A.}~\bibnamefont {Blais}}, \bibinfo {author} {\bibfnamefont {M.~H.}\ \bibnamefont {Devoret}}, \bibinfo {author} {\bibfnamefont {S.~M.}\ \bibnamefont {Girvin}},\ and\ \bibinfo {author} {\bibfnamefont {R.~J.}\ \bibnamefont {Schoelkopf}},\ }\bibfield  {title} {\bibinfo {title} {{Charge-insensitive qubit design derived from the Cooper pair box}},\ }\href {https://doi.org/10.1103/physreva.76.042319} {\bibfield  {journal} {\bibinfo  {journal} {Phys. Rev. A}\ }\textbf {\bibinfo {volume} {76}},\ \bibinfo {pages} {042319} (\bibinfo {year} {2007})},\ \Eprint
  {https://arxiv.org/abs/cond-mat/0703002} {arXiv:cond-mat/0703002} \BibitemShut {NoStop}%
\bibitem [{\citenamefont {Hall}\ \emph {et~al.}(2021)\citenamefont {Hall}, \citenamefont {Roggero}, \citenamefont {Baroni},\ and\ \citenamefont {Carlson}}]{Hall:2021rbv}%
  \BibitemOpen
  \bibfield  {author} {\bibinfo {author} {\bibfnamefont {B.}~\bibnamefont {Hall}}, \bibinfo {author} {\bibfnamefont {A.}~\bibnamefont {Roggero}}, \bibinfo {author} {\bibfnamefont {A.}~\bibnamefont {Baroni}},\ and\ \bibinfo {author} {\bibfnamefont {J.}~\bibnamefont {Carlson}},\ }\bibfield  {title} {\bibinfo {title} {{Simulation of collective neutrino oscillations on a quantum computer}},\ }\href {https://doi.org/10.1103/PhysRevD.104.063009} {\bibfield  {journal} {\bibinfo  {journal} {Phys. Rev. D}\ }\textbf {\bibinfo {volume} {104}},\ \bibinfo {pages} {063009} (\bibinfo {year} {2021})},\ \Eprint {https://arxiv.org/abs/2102.12556} {arXiv:2102.12556 [quant-ph]} \BibitemShut {NoStop}%
\bibitem [{\citenamefont {Yeter-Aydeniz}\ \emph {et~al.}(2022)\citenamefont {Yeter-Aydeniz}, \citenamefont {Bangar}, \citenamefont {Siopsis},\ and\ \citenamefont {Pooser}}]{Yeter-Aydeniz:2021olz}%
  \BibitemOpen
  \bibfield  {author} {\bibinfo {author} {\bibfnamefont {K.}~\bibnamefont {Yeter-Aydeniz}}, \bibinfo {author} {\bibfnamefont {S.}~\bibnamefont {Bangar}}, \bibinfo {author} {\bibfnamefont {G.}~\bibnamefont {Siopsis}},\ and\ \bibinfo {author} {\bibfnamefont {R.~C.}\ \bibnamefont {Pooser}},\ }\bibfield  {title} {\bibinfo {title} {{Collective neutrino oscillations on a quantum computer}},\ }\href {https://doi.org/10.1007/s11128-021-03348-x} {\bibfield  {journal} {\bibinfo  {journal} {Quant. Inf. Proc.}\ }\textbf {\bibinfo {volume} {21}},\ \bibinfo {pages} {84} (\bibinfo {year} {2022})},\ \Eprint {https://arxiv.org/abs/2104.03273} {arXiv:2104.03273 [quant-ph]} \BibitemShut {NoStop}%
\bibitem [{\citenamefont {Illa}\ and\ \citenamefont {Savage}(2023)}]{Illa:2022zgu}%
  \BibitemOpen
  \bibfield  {author} {\bibinfo {author} {\bibfnamefont {M.}~\bibnamefont {Illa}}\ and\ \bibinfo {author} {\bibfnamefont {M.~J.}\ \bibnamefont {Savage}},\ }\bibfield  {title} {\bibinfo {title} {{Multi-Neutrino Entanglement and Correlations in Dense Neutrino Systems}},\ }\href {https://doi.org/10.1103/PhysRevLett.130.221003} {\bibfield  {journal} {\bibinfo  {journal} {Phys. Rev. Lett.}\ }\textbf {\bibinfo {volume} {130}},\ \bibinfo {pages} {221003} (\bibinfo {year} {2023})},\ \Eprint {https://arxiv.org/abs/2210.08656} {arXiv:2210.08656 [nucl-th]} \BibitemShut {NoStop}%
\bibitem [{\citenamefont {Amitrano}\ \emph {et~al.}(2023)\citenamefont {Amitrano}, \citenamefont {Roggero}, \citenamefont {Luchi}, \citenamefont {Turro}, \citenamefont {Vespucci},\ and\ \citenamefont {Pederiva}}]{Amitrano:2022yyn}%
  \BibitemOpen
  \bibfield  {author} {\bibinfo {author} {\bibfnamefont {V.}~\bibnamefont {Amitrano}}, \bibinfo {author} {\bibfnamefont {A.}~\bibnamefont {Roggero}}, \bibinfo {author} {\bibfnamefont {P.}~\bibnamefont {Luchi}}, \bibinfo {author} {\bibfnamefont {F.}~\bibnamefont {Turro}}, \bibinfo {author} {\bibfnamefont {L.}~\bibnamefont {Vespucci}},\ and\ \bibinfo {author} {\bibfnamefont {F.}~\bibnamefont {Pederiva}},\ }\bibfield  {title} {\bibinfo {title} {{Trapped-ion quantum simulation of collective neutrino oscillations}},\ }\href {https://doi.org/10.1103/PhysRevD.107.023007} {\bibfield  {journal} {\bibinfo  {journal} {Phys. Rev. D}\ }\textbf {\bibinfo {volume} {107}},\ \bibinfo {pages} {023007} (\bibinfo {year} {2023})},\ \Eprint {https://arxiv.org/abs/2207.03189} {arXiv:2207.03189 [quant-ph]} \BibitemShut {NoStop}%
\bibitem [{\citenamefont {Siwach}\ \emph {et~al.}(2023)\citenamefont {Siwach}, \citenamefont {Harrison},\ and\ \citenamefont {Balantekin}}]{Siwach:2023wzy}%
  \BibitemOpen
  \bibfield  {author} {\bibinfo {author} {\bibfnamefont {P.}~\bibnamefont {Siwach}}, \bibinfo {author} {\bibfnamefont {K.}~\bibnamefont {Harrison}},\ and\ \bibinfo {author} {\bibfnamefont {A.~B.}\ \bibnamefont {Balantekin}},\ }\bibfield  {title} {\bibinfo {title} {{Collective neutrino oscillations on a quantum computer with hybrid quantum-classical algorithm}},\ }\href {https://doi.org/10.1103/PhysRevD.108.083039} {\bibfield  {journal} {\bibinfo  {journal} {Phys. Rev. D}\ }\textbf {\bibinfo {volume} {108}},\ \bibinfo {pages} {083039} (\bibinfo {year} {2023})},\ \Eprint {https://arxiv.org/abs/2308.09123} {arXiv:2308.09123 [quant-ph]} \BibitemShut {NoStop}%
\bibitem [{\citenamefont {Turro}\ \emph {et~al.}(2025)\citenamefont {Turro}, \citenamefont {Chernyshev}, \citenamefont {Bhaskar},\ and\ \citenamefont {Illa}}]{Turro:2024shh}%
  \BibitemOpen
  \bibfield  {author} {\bibinfo {author} {\bibfnamefont {F.}~\bibnamefont {Turro}}, \bibinfo {author} {\bibfnamefont {I.~A.}\ \bibnamefont {Chernyshev}}, \bibinfo {author} {\bibfnamefont {R.}~\bibnamefont {Bhaskar}},\ and\ \bibinfo {author} {\bibfnamefont {M.}~\bibnamefont {Illa}},\ }\bibfield  {title} {\bibinfo {title} {{Qutrit and qubit circuits for three-flavor collective neutrino oscillations}},\ }\href {https://doi.org/10.1103/PhysRevD.111.043038} {\bibfield  {journal} {\bibinfo  {journal} {Phys. Rev. D}\ }\textbf {\bibinfo {volume} {111}},\ \bibinfo {pages} {043038} (\bibinfo {year} {2025})},\ \Eprint {https://arxiv.org/abs/2407.13914} {arXiv:2407.13914 [quant-ph]} \BibitemShut {NoStop}%
\bibitem [{\citenamefont {Spagnoli}\ \emph {et~al.}(2025)\citenamefont {Spagnoli} \emph {et~al.}}]{Spagnoli:2025etu}%
  \BibitemOpen
  \bibfield  {author} {\bibinfo {author} {\bibfnamefont {L.}~\bibnamefont {Spagnoli}} \emph {et~al.},\ }\bibfield  {title} {\bibinfo {title} {{Collective neutrino oscillations in three flavors on qubit and qutrit processors}},\ }\href {https://doi.org/10.1103/gjr1-lf8s} {\bibfield  {journal} {\bibinfo  {journal} {Phys. Rev. D}\ }\textbf {\bibinfo {volume} {111}},\ \bibinfo {pages} {103054} (\bibinfo {year} {2025})},\ \Eprint {https://arxiv.org/abs/2503.00607} {arXiv:2503.00607 [quant-ph]} \BibitemShut {NoStop}%
\bibitem [{\citenamefont {Preskill}(2018)}]{Preskill:2018jim}%
  \BibitemOpen
  \bibfield  {author} {\bibinfo {author} {\bibfnamefont {J.}~\bibnamefont {Preskill}},\ }\bibfield  {title} {\bibinfo {title} {{Quantum Computing in the NISQ era and beyond}},\ }\href {https://doi.org/10.22331/q-2018-08-06-79} {\bibfield  {journal} {\bibinfo  {journal} {Quantum}\ }\textbf {\bibinfo {volume} {2}},\ \bibinfo {pages} {79} (\bibinfo {year} {2018})},\ \Eprint {https://arxiv.org/abs/1801.00862} {arXiv:1801.00862 [quant-ph]} \BibitemShut {NoStop}%
\bibitem [{\citenamefont {Vatan}\ and\ \citenamefont {Williams}(2004)}]{Vatan:2004nmz}%
  \BibitemOpen
  \bibfield  {author} {\bibinfo {author} {\bibfnamefont {F.}~\bibnamefont {Vatan}}\ and\ \bibinfo {author} {\bibfnamefont {C.}~\bibnamefont {Williams}},\ }\bibfield  {title} {\bibinfo {title} {{Optimal quantum circuits for general two-qubit gates}},\ }\href {https://doi.org/10.1103/PhysRevA.69.032315} {\bibfield  {journal} {\bibinfo  {journal} {Phys. Rev. A}\ }\textbf {\bibinfo {volume} {69}},\ \bibinfo {pages} {032315} (\bibinfo {year} {2004})}\BibitemShut {NoStop}%
\bibitem [{\citenamefont {Shende}\ \emph {et~al.}(2004)\citenamefont {Shende}, \citenamefont {Markov},\ and\ \citenamefont {Bullock}}]{Shende:2004gqq}%
  \BibitemOpen
  \bibfield  {author} {\bibinfo {author} {\bibfnamefont {V.~V.}\ \bibnamefont {Shende}}, \bibinfo {author} {\bibfnamefont {I.~L.}\ \bibnamefont {Markov}},\ and\ \bibinfo {author} {\bibfnamefont {S.~S.}\ \bibnamefont {Bullock}},\ }\bibfield  {title} {\bibinfo {title} {{Minimal universal two-qubit controlled-NOT-based circuits}},\ }\href {https://doi.org/10.1103/PhysRevA.69.062321} {\bibfield  {journal} {\bibinfo  {journal} {Phys. Rev. A}\ }\textbf {\bibinfo {volume} {69}},\ \bibinfo {pages} {062321} (\bibinfo {year} {2004})}\BibitemShut {NoStop}%
\bibitem [{\citenamefont {Jang}\ \emph {et~al.}(2021)\citenamefont {Jang}, \citenamefont {Terashi}, \citenamefont {Saito}, \citenamefont {Bauer}, \citenamefont {Nachman}, \citenamefont {Iiyama}, \citenamefont {Kishimoto}, \citenamefont {Okubo}, \citenamefont {Sawada},\ and\ \citenamefont {Tanaka}}]{Jang:2021ary}%
  \BibitemOpen
  \bibfield  {author} {\bibinfo {author} {\bibfnamefont {W.}~\bibnamefont {Jang}}, \bibinfo {author} {\bibfnamefont {K.}~\bibnamefont {Terashi}}, \bibinfo {author} {\bibfnamefont {M.}~\bibnamefont {Saito}}, \bibinfo {author} {\bibfnamefont {C.~W.}\ \bibnamefont {Bauer}}, \bibinfo {author} {\bibfnamefont {B.}~\bibnamefont {Nachman}}, \bibinfo {author} {\bibfnamefont {Y.}~\bibnamefont {Iiyama}}, \bibinfo {author} {\bibfnamefont {T.}~\bibnamefont {Kishimoto}}, \bibinfo {author} {\bibfnamefont {R.}~\bibnamefont {Okubo}}, \bibinfo {author} {\bibfnamefont {R.}~\bibnamefont {Sawada}},\ and\ \bibinfo {author} {\bibfnamefont {J.}~\bibnamefont {Tanaka}},\ }\bibfield  {title} {\bibinfo {title} {{Quantum Gate Pattern Recognition and Circuit Optimization for Scientific Applications}},\ }\href {https://doi.org/10.1051/epjconf/202125103023} {\bibfield  {journal} {\bibinfo  {journal} {EPJ Web Conf.}\ }\textbf {\bibinfo {volume} {251}},\ \bibinfo {pages} {03023} (\bibinfo {year} {2021})},\ \Eprint
  {https://arxiv.org/abs/2102.10008} {arXiv:2102.10008 [quant-ph]} \BibitemShut {NoStop}%
\bibitem [{\citenamefont {Zulehner}\ \emph {et~al.}(2019)\citenamefont {Zulehner}, \citenamefont {Paler},\ and\ \citenamefont {Wille}}]{Zulehner:2019lkn}%
  \BibitemOpen
  \bibfield  {author} {\bibinfo {author} {\bibfnamefont {A.}~\bibnamefont {Zulehner}}, \bibinfo {author} {\bibfnamefont {A.}~\bibnamefont {Paler}},\ and\ \bibinfo {author} {\bibfnamefont {R.}~\bibnamefont {Wille}},\ }\bibfield  {title} {\bibinfo {title} {{An Efficient Methodology for Mapping Quantum Circuits to the IBM QX Architectures}},\ }\href {https://doi.org/10.1109/TCAD.2018.2846658} {\bibfield  {journal} {\bibinfo  {journal} {IEEE Trans. Comput. Aided Design Integr. Circuits Syst.}\ }\textbf {\bibinfo {volume} {38}},\ \bibinfo {pages} {1226} (\bibinfo {year} {2019})}\BibitemShut {NoStop}%
\bibitem [{\citenamefont {Wille}\ \emph {et~al.}(2019)\citenamefont {Wille}, \citenamefont {Burgholzer},\ and\ \citenamefont {Zulehner}}]{Wille:2019lxk}%
  \BibitemOpen
  \bibfield  {author} {\bibinfo {author} {\bibfnamefont {R.}~\bibnamefont {Wille}}, \bibinfo {author} {\bibfnamefont {L.}~\bibnamefont {Burgholzer}},\ and\ \bibinfo {author} {\bibfnamefont {A.}~\bibnamefont {Zulehner}},\ }\bibfield  {title} {\bibinfo {title} {{Mapping Quantum Circuits to IBM QX Architectures Using the Minimal Number of SWAP and H Operations}},\ }\href@noop {} {\  (\bibinfo {year} {2019})},\ \Eprint {https://arxiv.org/abs/1907.02026} {arXiv:1907.02026 [quant-ph]} \BibitemShut {NoStop}%
\bibitem [{\citenamefont {Lin}\ \emph {et~al.}(2022)\citenamefont {Lin}, \citenamefont {Tan}, \citenamefont {Niu}, \citenamefont {Kimko},\ and\ \citenamefont {Cong}}]{Lin:2022cyg}%
  \BibitemOpen
  \bibfield  {author} {\bibinfo {author} {\bibfnamefont {W.-H.}\ \bibnamefont {Lin}}, \bibinfo {author} {\bibfnamefont {B.}~\bibnamefont {Tan}}, \bibinfo {author} {\bibfnamefont {M.~Y.}\ \bibnamefont {Niu}}, \bibinfo {author} {\bibfnamefont {J.}~\bibnamefont {Kimko}},\ and\ \bibinfo {author} {\bibfnamefont {J.}~\bibnamefont {Cong}},\ }\bibfield  {title} {\bibinfo {title} {{Domain-Specific Quantum Architecture Optimization}},\ }\href@noop {} {\  (\bibinfo {year} {2022})},\ \Eprint {https://arxiv.org/abs/2207.14482} {arXiv:2207.14482 [cs.AR]} \BibitemShut {NoStop}%
\bibitem [{\citenamefont {DeCross}\ \emph {et~al.}(2023)\citenamefont {DeCross}, \citenamefont {Chertkov}, \citenamefont {Kohagen},\ and\ \citenamefont {Foss-Feig}}]{DeCross:2022kuu}%
  \BibitemOpen
  \bibfield  {author} {\bibinfo {author} {\bibfnamefont {M.}~\bibnamefont {DeCross}}, \bibinfo {author} {\bibfnamefont {E.}~\bibnamefont {Chertkov}}, \bibinfo {author} {\bibfnamefont {M.}~\bibnamefont {Kohagen}},\ and\ \bibinfo {author} {\bibfnamefont {M.}~\bibnamefont {Foss-Feig}},\ }\bibfield  {title} {\bibinfo {title} {{Qubit-Reuse Compilation with Mid-Circuit Measurement and Reset}},\ }\href {https://doi.org/10.1103/PhysRevX.13.041057} {\bibfield  {journal} {\bibinfo  {journal} {Phys. Rev. X}\ }\textbf {\bibinfo {volume} {13}},\ \bibinfo {pages} {041057} (\bibinfo {year} {2023})},\ \Eprint {https://arxiv.org/abs/2210.08039} {arXiv:2210.08039 [quant-ph]} \BibitemShut {NoStop}%
\bibitem [{\citenamefont {Fischer}\ \emph {et~al.}(2023)\citenamefont {Fischer}, \citenamefont {Chiesa}, \citenamefont {Tacchino}, \citenamefont {Egger}, \citenamefont {Carretta},\ and\ \citenamefont {Tavernelli}}]{Fischer:2022dbr}%
  \BibitemOpen
  \bibfield  {author} {\bibinfo {author} {\bibfnamefont {L.~E.}\ \bibnamefont {Fischer}}, \bibinfo {author} {\bibfnamefont {A.}~\bibnamefont {Chiesa}}, \bibinfo {author} {\bibfnamefont {F.}~\bibnamefont {Tacchino}}, \bibinfo {author} {\bibfnamefont {D.~J.}\ \bibnamefont {Egger}}, \bibinfo {author} {\bibfnamefont {S.}~\bibnamefont {Carretta}},\ and\ \bibinfo {author} {\bibfnamefont {I.}~\bibnamefont {Tavernelli}},\ }\bibfield  {title} {\bibinfo {title} {{Universal Qudit Gate Synthesis for Transmons}},\ }\href {https://doi.org/10.1103/PRXQuantum.4.030327} {\bibfield  {journal} {\bibinfo  {journal} {PRX Quantum}\ }\textbf {\bibinfo {volume} {4}},\ \bibinfo {pages} {030327} (\bibinfo {year} {2023})},\ \Eprint {https://arxiv.org/abs/2212.04496} {arXiv:2212.04496 [quant-ph]} \BibitemShut {NoStop}%
\bibitem [{\citenamefont {Kiktenko}\ \emph {et~al.}(2025)\citenamefont {Kiktenko}, \citenamefont {Nikolaeva},\ and\ \citenamefont {Fedorov}}]{Kiktenko:2023ytz}%
  \BibitemOpen
  \bibfield  {author} {\bibinfo {author} {\bibfnamefont {E.~O.}\ \bibnamefont {Kiktenko}}, \bibinfo {author} {\bibfnamefont {A.~S.}\ \bibnamefont {Nikolaeva}},\ and\ \bibinfo {author} {\bibfnamefont {A.~K.}\ \bibnamefont {Fedorov}},\ }\bibfield  {title} {\bibinfo {title} {{Colloquium: Qudits for decomposing multiqubit gates and realizing quantum algorithms}},\ }\href {https://doi.org/10.1103/RevModPhys.97.021003} {\bibfield  {journal} {\bibinfo  {journal} {Rev. Mod. Phys.}\ }\textbf {\bibinfo {volume} {97}},\ \bibinfo {pages} {021003} (\bibinfo {year} {2025})},\ \Eprint {https://arxiv.org/abs/2311.12003} {arXiv:2311.12003 [quant-ph]} \BibitemShut {NoStop}%
\bibitem [{\citenamefont {Vezvaee}\ \emph {et~al.}(2025)\citenamefont {Vezvaee}, \citenamefont {Tripathi}, \citenamefont {Kowsari}, \citenamefont {Levenson-Falk},\ and\ \citenamefont {Lidar}}]{Vezvaee:2024ywq}%
  \BibitemOpen
  \bibfield  {author} {\bibinfo {author} {\bibfnamefont {A.}~\bibnamefont {Vezvaee}}, \bibinfo {author} {\bibfnamefont {V.}~\bibnamefont {Tripathi}}, \bibinfo {author} {\bibfnamefont {D.}~\bibnamefont {Kowsari}}, \bibinfo {author} {\bibfnamefont {E.}~\bibnamefont {Levenson-Falk}},\ and\ \bibinfo {author} {\bibfnamefont {D.~A.}\ \bibnamefont {Lidar}},\ }\bibfield  {title} {\bibinfo {title} {{Virtual-Z Gates and Symmetric Gate Compilation}},\ }\href {https://doi.org/10.1103/PRXQuantum.6.020348} {\bibfield  {journal} {\bibinfo  {journal} {PRX Quantum}\ }\textbf {\bibinfo {volume} {6}},\ \bibinfo {pages} {020348} (\bibinfo {year} {2025})},\ \Eprint {https://arxiv.org/abs/2407.14782} {arXiv:2407.14782 [quant-ph]} \BibitemShut {NoStop}%
\bibitem [{\citenamefont {McKay}\ \emph {et~al.}(2017)\citenamefont {McKay}, \citenamefont {Wood}, \citenamefont {Sheldon}, \citenamefont {Chow},\ and\ \citenamefont {Gambetta}}]{McKay:2017rej}%
  \BibitemOpen
  \bibfield  {author} {\bibinfo {author} {\bibfnamefont {D.~C.}\ \bibnamefont {McKay}}, \bibinfo {author} {\bibfnamefont {C.~J.}\ \bibnamefont {Wood}}, \bibinfo {author} {\bibfnamefont {S.}~\bibnamefont {Sheldon}}, \bibinfo {author} {\bibfnamefont {J.~M.}\ \bibnamefont {Chow}},\ and\ \bibinfo {author} {\bibfnamefont {J.~M.}\ \bibnamefont {Gambetta}},\ }\bibfield  {title} {\bibinfo {title} {{Efficient Z gates for quantum computing}},\ }\href {https://doi.org/10.1103/PhysRevA.96.022330} {\bibfield  {journal} {\bibinfo  {journal} {Phys. Rev. A}\ }\textbf {\bibinfo {volume} {96}},\ \bibinfo {pages} {022330} (\bibinfo {year} {2017})}\BibitemShut {NoStop}%
\bibitem [{\citenamefont {Akhmedov}(2000)}]{Akhmedov:1999ty}%
  \BibitemOpen
  \bibfield  {author} {\bibinfo {author} {\bibfnamefont {E.~K.}\ \bibnamefont {Akhmedov}},\ }\bibfield  {title} {\bibinfo {title} {{Parametric resonance in neutrino oscillations in matter}},\ }\href {https://doi.org/10.1007/s12043-000-0006-4} {\bibfield  {journal} {\bibinfo  {journal} {Pramana}\ }\textbf {\bibinfo {volume} {54}},\ \bibinfo {pages} {47} (\bibinfo {year} {2000})},\ \Eprint {https://arxiv.org/abs/hep-ph/9907435} {arXiv:hep-ph/9907435} \BibitemShut {NoStop}%
\bibitem [{\citenamefont {Akhmedov}(1999)}]{Akhmedov:1998ui}%
  \BibitemOpen
  \bibfield  {author} {\bibinfo {author} {\bibfnamefont {E.~K.}\ \bibnamefont {Akhmedov}},\ }\bibfield  {title} {\bibinfo {title} {{Parametric resonance of neutrino oscillations and passage of solar and atmospheric neutrinos through the earth}},\ }\href {https://doi.org/10.1016/S0550-3213(98)00723-8} {\bibfield  {journal} {\bibinfo  {journal} {Nucl. Phys. B}\ }\textbf {\bibinfo {volume} {538}},\ \bibinfo {pages} {25} (\bibinfo {year} {1999})},\ \Eprint {https://arxiv.org/abs/hep-ph/9805272} {arXiv:hep-ph/9805272} \BibitemShut {NoStop}%
\bibitem [{\citenamefont {Maltoni}\ and\ \citenamefont {Smirnov}(2016)}]{Maltoni:2015kca}%
  \BibitemOpen
  \bibfield  {author} {\bibinfo {author} {\bibfnamefont {M.}~\bibnamefont {Maltoni}}\ and\ \bibinfo {author} {\bibfnamefont {A.~{\relax Yu}.}\ \bibnamefont {Smirnov}},\ }\bibfield  {title} {\bibinfo {title} {{Solar neutrinos and neutrino physics}},\ }\href {https://doi.org/10.1140/epja/i2016-16087-0} {\bibfield  {journal} {\bibinfo  {journal} {Eur. Phys. J.}\ }\textbf {\bibinfo {volume} {A52}},\ \bibinfo {pages} {87} (\bibinfo {year} {2016})},\ \Eprint {https://arxiv.org/abs/1507.05287} {arXiv:1507.05287 [hep-ph]} \BibitemShut {NoStop}%
\bibitem [{\citenamefont {Javadi-Abhari}\ \emph {et~al.}(2024)\citenamefont {Javadi-Abhari}, \citenamefont {Treinish}, \citenamefont {Krsulich}, \citenamefont {Wood}, \citenamefont {Lishman}, \citenamefont {Gacon}, \citenamefont {Martiel}, \citenamefont {Nation}, \citenamefont {Bishop}, \citenamefont {Cross}, \citenamefont {Johnson},\ and\ \citenamefont {Gambetta}}]{qiskit2024}%
  \BibitemOpen
  \bibfield  {author} {\bibinfo {author} {\bibfnamefont {A.}~\bibnamefont {Javadi-Abhari}}, \bibinfo {author} {\bibfnamefont {M.}~\bibnamefont {Treinish}}, \bibinfo {author} {\bibfnamefont {K.}~\bibnamefont {Krsulich}}, \bibinfo {author} {\bibfnamefont {C.~J.}\ \bibnamefont {Wood}}, \bibinfo {author} {\bibfnamefont {J.}~\bibnamefont {Lishman}}, \bibinfo {author} {\bibfnamefont {J.}~\bibnamefont {Gacon}}, \bibinfo {author} {\bibfnamefont {S.}~\bibnamefont {Martiel}}, \bibinfo {author} {\bibfnamefont {P.~D.}\ \bibnamefont {Nation}}, \bibinfo {author} {\bibfnamefont {L.~S.}\ \bibnamefont {Bishop}}, \bibinfo {author} {\bibfnamefont {A.~W.}\ \bibnamefont {Cross}}, \bibinfo {author} {\bibfnamefont {B.~R.}\ \bibnamefont {Johnson}},\ and\ \bibinfo {author} {\bibfnamefont {J.~M.}\ \bibnamefont {Gambetta}},\ }\href {https://doi.org/10.48550/arXiv.2405.08810} {\bibinfo {title} {Quantum computing with {Q}iskit}} (\bibinfo {year} {2024}),\ \Eprint {https://arxiv.org/abs/2405.08810} {arXiv:2405.08810 [quant-ph]}
  \BibitemShut {NoStop}%
\bibitem [{\citenamefont {Hosaka}\ \emph {et~al.}(2006)\citenamefont {Hosaka} \emph {et~al.}}]{Super-Kamiokande:2006jvq}%
  \BibitemOpen
  \bibfield  {author} {\bibinfo {author} {\bibfnamefont {J.}~\bibnamefont {Hosaka}} \emph {et~al.} (\bibinfo {collaboration} {Super-Kamiokande}),\ }\bibfield  {title} {\bibinfo {title} {{Three flavor neutrino oscillation analysis of atmospheric neutrinos in Super-Kamiokande}},\ }\href {https://doi.org/10.1103/PhysRevD.74.032002} {\bibfield  {journal} {\bibinfo  {journal} {Phys. Rev. D}\ }\textbf {\bibinfo {volume} {74}},\ \bibinfo {pages} {032002} (\bibinfo {year} {2006})},\ \Eprint {https://arxiv.org/abs/hep-ex/0604011} {arXiv:hep-ex/0604011} \BibitemShut {NoStop}%
\bibitem [{\citenamefont {Vidal}\ and\ \citenamefont {Dawson}(2004)}]{Vidal:2004akd}%
  \BibitemOpen
  \bibfield  {author} {\bibinfo {author} {\bibfnamefont {G.}~\bibnamefont {Vidal}}\ and\ \bibinfo {author} {\bibfnamefont {C.~M.}\ \bibnamefont {Dawson}},\ }\bibfield  {title} {\bibinfo {title} {{Universal quantum circuit for two-qubit transformations with three controlled-NOT gates}},\ }\href {https://doi.org/10.1103/PhysRevA.69.010301} {\bibfield  {journal} {\bibinfo  {journal} {Phys. Rev. A}\ }\textbf {\bibinfo {volume} {69}},\ \bibinfo {pages} {010301} (\bibinfo {year} {2004})}\BibitemShut {NoStop}%
\bibitem [{\citenamefont {Nielsen}\ and\ \citenamefont {Chuang}(2012)}]{Nielsen:2012yss}%
  \BibitemOpen
  \bibfield  {author} {\bibinfo {author} {\bibfnamefont {M.~A.}\ \bibnamefont {Nielsen}}\ and\ \bibinfo {author} {\bibfnamefont {I.~L.}\ \bibnamefont {Chuang}},\ }\href {https://doi.org/10.1017/cbo9780511976667} {\emph {\bibinfo {title} {{Quantum Computation and Quantum Information}}}}\ (\bibinfo  {publisher} {Cambridge University Press},\ \bibinfo {year} {2012})\BibitemShut {NoStop}%
\bibitem [{\citenamefont {Giunti}\ and\ \citenamefont {Studenikin}(2015)}]{Giunti:2014ixa}%
  \BibitemOpen
  \bibfield  {author} {\bibinfo {author} {\bibfnamefont {C.}~\bibnamefont {Giunti}}\ and\ \bibinfo {author} {\bibfnamefont {A.}~\bibnamefont {Studenikin}},\ }\bibfield  {title} {\bibinfo {title} {{Neutrino electromagnetic interactions: a window to new physics}},\ }\href {https://doi.org/10.1103/RevModPhys.87.531} {\bibfield  {journal} {\bibinfo  {journal} {Rev. Mod. Phys.}\ }\textbf {\bibinfo {volume} {87}},\ \bibinfo {pages} {531} (\bibinfo {year} {2015})},\ \Eprint {https://arxiv.org/abs/1403.6344} {arXiv:1403.6344 [hep-ph]} \BibitemShut {NoStop}%
\bibitem [{\citenamefont {Joshi}\ and\ \citenamefont {Jain}(2017)}]{Joshi:2017vpi}%
  \BibitemOpen
  \bibfield  {author} {\bibinfo {author} {\bibfnamefont {S.}~\bibnamefont {Joshi}}\ and\ \bibinfo {author} {\bibfnamefont {S.~R.}\ \bibnamefont {Jain}},\ }\bibfield  {title} {\bibinfo {title} {{Noncyclic geometric phases and helicity transitions for neutrino oscillations in a magnetic field}},\ }\href {https://doi.org/10.1103/PhysRevD.96.096004} {\bibfield  {journal} {\bibinfo  {journal} {Phys. Rev. D}\ }\textbf {\bibinfo {volume} {96}},\ \bibinfo {pages} {096004} (\bibinfo {year} {2017})},\ \Eprint {https://arxiv.org/abs/1703.05027} {arXiv:1703.05027 [hep-ph]} \BibitemShut {NoStop}%
\bibitem [{\citenamefont {Benatti}\ and\ \citenamefont {Floreanini}(2000)}]{Benatti:2000ph}%
  \BibitemOpen
  \bibfield  {author} {\bibinfo {author} {\bibfnamefont {F.}~\bibnamefont {Benatti}}\ and\ \bibinfo {author} {\bibfnamefont {R.}~\bibnamefont {Floreanini}},\ }\bibfield  {title} {\bibinfo {title} {{Open system approach to neutrino oscillations}},\ }\href {https://doi.org/10.1088/1126-6708/2000/02/032} {\bibfield  {journal} {\bibinfo  {journal} {JHEP}\ }\textbf {\bibinfo {volume} {02}},\ \bibinfo {pages} {032}},\ \Eprint {https://arxiv.org/abs/hep-ph/0002221} {arXiv:hep-ph/0002221} \BibitemShut {NoStop}%
\bibitem [{\citenamefont {Oliveira}\ and\ \citenamefont {Guzzo}(2010)}]{Oliveira:2010zzd}%
  \BibitemOpen
  \bibfield  {author} {\bibinfo {author} {\bibfnamefont {R.~L.~N.}\ \bibnamefont {Oliveira}}\ and\ \bibinfo {author} {\bibfnamefont {M.~M.}\ \bibnamefont {Guzzo}},\ }\bibfield  {title} {\bibinfo {title} {{Quantum dissipation in vacuum neutrino oscillation}},\ }\href {https://doi.org/10.1140/epjc/s10052-010-1388-1} {\bibfield  {journal} {\bibinfo  {journal} {Eur. Phys. J. C}\ }\textbf {\bibinfo {volume} {69}},\ \bibinfo {pages} {493} (\bibinfo {year} {2010})}\BibitemShut {NoStop}%
\bibitem [{\citenamefont {Ohlsson}(2013)}]{Ohlsson:2012kf}%
  \BibitemOpen
  \bibfield  {author} {\bibinfo {author} {\bibfnamefont {T.}~\bibnamefont {Ohlsson}},\ }\bibfield  {title} {\bibinfo {title} {{Status of non-standard neutrino interactions}},\ }\href {https://doi.org/10.1088/0034-4885/76/4/044201} {\bibfield  {journal} {\bibinfo  {journal} {Rept. Prog. Phys.}\ }\textbf {\bibinfo {volume} {76}},\ \bibinfo {pages} {044201} (\bibinfo {year} {2013})},\ \Eprint {https://arxiv.org/abs/1209.2710} {arXiv:1209.2710 [hep-ph]} \BibitemShut {NoStop}%
\end{thebibliography}%

\end{document}